\documentclass[
aps,
prl,
reprint,
superscriptaddress,
amsmath,
amssymb,
showpacs
]{revtex4-2}

\usepackage{graphicx}
\usepackage{multirow}
\usepackage{color}

\newcommand{\DD}[2]{\frac{\text{d}#1}{\text{d}#2}}

\newcommand{\Bunit}{A/m$^2$/rad$^2$}

\begin{document}

% Use the \preprint command to place your local instit{}utional report
%\preprint{}

\title{Ultra-Bright Electron Bunch Injection in a Plasma Wakefield Driven by a Superluminal Flying Focus Electron Beam}

\author{F. Li}
\email{lifei11@ucla.edu}
\affiliation{Department of Electrical Engineering, University of California Los Angeles, Los Angeles, California 90095, USA}
\author{T. N. Dalichaouch}
\affiliation{Department of Physics and Astronomy, University of California Los Angeles, Los Angeles, California 90095, USA}
\author{J. R. Pierce}
\affiliation{Department of Physics and Astronomy, University of California Los Angeles, Los Angeles, California 90095, USA}
\author{X. Xu}
\affiliation{SLAC National Accelerator Laboratory, Menlo Park, California 94025, USA}
\author{F. S. Tsung}
\affiliation{Department of Physics and Astronomy, University of California Los Angeles, Los Angeles, California 90095, USA}
\author{W. Lu}
\affiliation{Department of Engineering Physics, Tsinghua University, Beijing 100084, China}
\author{C. Joshi}
\affiliation{Department of Electrical Engineering, University of California Los Angeles, Los Angeles, California 90095, USA}
\author{W. B. Mori}
\email{mori@physics.ucla.edu}
\affiliation{Department of Electrical Engineering, University of California Los Angeles, Los Angeles, California 90095, USA}
\affiliation{Department of Physics and Astronomy, University of California Los Angeles, Los Angeles, California 90095, USA}

\date{\today}

\begin{abstract}

We propose a new method for self-injection of high-quality electron bunches in the plasma wakefield structure in the blowout regime utilizing a ``flying focus" produced by a drive beam with an energy chirp. In a flying focus the speed of the density centroid of the drive bunch can be superluminal or subluminal by utilizing the chromatic dependence of the focusing optics. We first derive the focal velocity and the characteristic length of the focal spot in terms of the focal length and an energy chirp. We then demonstrate using multidimensional particle-in-cell simulations that a wake driven by a superluminally propagating flying focus of an electron beam can generate GeV-level electron bunches with ultralow normalized slice emittance ($\sim$30 nm rad), high current ($\sim$17 kA), low slice energy spread ($\sim$0.1\%) and  therefore high normalized brightness ($>10^{19}$ \Bunit) in a plasma of density $\sim10^{19}$ cm$^{-3}$. The injection process is highly controllable and tunable by changing the focal velocity and shaping the drive beam current.
Near-term experiments at FACET II where the capabilities to generate tens of kA, <10 fs drivers are planned, could potentially produce beams with brightness near $10^{20}$ \Bunit.

\end{abstract}

% \pacs{52.38.Kd, 41.75.Jv, 52.35.Mw}

%\keywords{}

\maketitle

Plasma–based accelerators (PBA) driven by either an intense laser pulse (LWFA) or a charged particle beam (PWFA) \cite{Joshi2003}, can sustain ultrahigh acceleration gradients $\sim$100GV/m and have the potential to produce high-quality electron beams. Numerous milestones in the PBA have been attained in the past two decades \cite{Joshi2020}. In the near term, a combination of high gradient and high beam quality may lead to a compact x-ray free-electron laser (XFEL) \cite{Wang2021,Xu2021} and new photon science applications. Electron beams needed to drive XFELs have stringent requirements on normalized beam emittance, energy spread and brightness \cite{Pellegrini2016}. Controllable injection in the plasma wake is a critical physical process that can determine the eventual beam quality. Various synchronized injection schemes, including field ionization injection \cite{Chen2006,Oz2007,Pak2010,Hidding2012,Li2013,MartinezDeLaOssa2013,Xu2014} and expanding plasma wakefields induced by either density tailoring \cite{Bulanov1998,Suk2001,Buck2013,Geddes2008,Xu2017} or drive beam evolution \cite{Kalmykov2009,Lehe2013,Dalichaouch2020}, have been proposed and in some cases studied in experiments. Simulations have shown that some schemes can produce the beam parameters needed for XFEL.

In this Letter, we propose and demonstrate using PIC simulations a new electron injection scheme in the three-dimensional nonlinear blowout regime \cite{Rosenzweig1991,Pukhov2002,Lu2006a,Lu2006b} of a PWFA. The injection process is triggered by a drive beam whose density centroid [``flying focus'' (FF)] moves superluminally. This may seem counterintuitive as injection occurs when the phase velocity of the wake decreases sufficiently such that electrons comove at the phase velocity of the wake. However, as elaborated below, when the FF is superluminal and the most dense (smallest spot size) part of the drive beam excites a nonlinear wake, an increasing amount of charge is confined within the ion channel leading to a backward expanding wake (in the comoving frame with the beam) as the density peak moves forward which effectively reduces the phase velocity at the rear of the wake. The proposed scheme is highly controllable and capable of generating GeV-level electron bunches with normalized emittances $\sim$10s nm, slice energy spreads $\sim$0.1\%, and normalized brightness $>10^{19}$ \Bunit, which is orders of magnitude higher than those of existing beams at state-of-the-art XFELs based on conventional accelerators.
Thus, this PBA based scheme not only provides the possibility of replacing the conventional injector and accelerator in an XFEL, but it could also provide a path for significantly boosting the brightness of an existing beam leading to compact, cost-effective XFELs since the saturation gain length and thus the required undulator length can be greatly reduced.
Unlike density-downramp injection that relies on tailoring the plasma density and some ionization schemes which need synchronization of laser pulses with the drive pulse, the proposed scheme relies on a simpler experimental setup --- a uniform plasma and a single drive beam. While both the evolving beam \cite{Dalichaouch2020} and FF schemes rely on the focusing optics to trigger injection, the FF scheme may provide better tunability, as well as controllability and stable acceleration after injection. Furthermore, typical beams produced at facilities such as FACET already have quasilinear energy chirps (at least on a significant portion of the beam) so the physics of an FF needs to be considered in general. 

Recently, the optical FF concept \cite{Froula2018,Palastro2018,Kondakci2019,Palastro2020,Yessenov2020,Caizergues2020} has been developed to provide customized spatiotemporal control over the intensity of focused laser beams. It has been proposed to use such pulses to overcome the dephasing that arises in LWFAs \cite{Palastro2020,Caizergues2020}.
In this Letter, we propose the use of the FF formed by a charged particle beam to trigger self-injection and accelerate the injected beam beam in a PWFA. The FF is formed from a charged particle beam with a correlated energy spread (energy chirp) focused by magnetic or plasma lenses. Since the focal length is proportional to the particle energy, different slices of a beam with an energy chirp will be focused to different positions due to chromatic aberrations. For a positive- (negative-) chirped beam, the slices approaching the beam head will come to focus earlier (latter), resulting in a sub- (super-) luminal FF. We note that ultrashort electron bunches with residual negative or positive chirps have been routinely produced at the Final Focus Test Beam experimental facility \cite{Hogan2005,Litos2014}.

\begin{figure}[htbp]
\includegraphics[width=8.6cm]{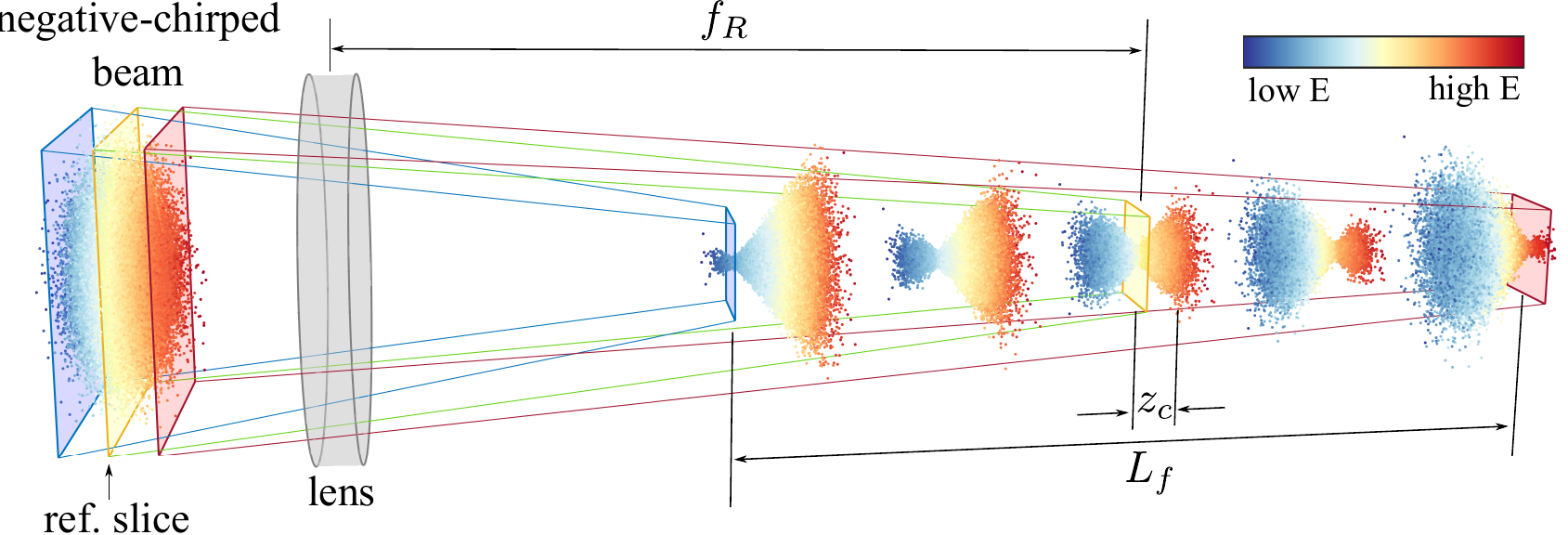}
\caption{
A schematic of generating a superluminal FF using a negative-chirped beam.
}
\label{fig:schematic_ff}
\end{figure}

Figure \ref{fig:schematic_ff} illustrates how to generate a superluminal FF.
An FF beam is characterized by the velocity of the density peak, the effective pulse length, and the effective diffraction length as illustrated in Fig. \ref{fig:schematic_ff}.

The normalized focal velocity $\beta_f$ is given by (see the Supplemental Material)
\begin{equation}
\label{eq:focal_velocity}
  \beta_f \simeq 1 - \frac{p_R}{f_R}\left(\DD{p}{\xi}\right)^{-1} 
  + \frac{1}{2\gamma^2}\left[1-\frac{3p_R}{f_R}\left(\DD{p}{\xi}\right)^{-1}\right],
\end{equation}
where $p$ and $p_R$ are the momenta of an arbitrary particle and the reference particle respectively, $f_R$ is the focal length of the reference particle, and $\xi\equiv ct-z$. For $\gamma^2\gg(f_R/p_R)|dp/d\xi|$, the $1/\gamma^2$ term arising from the interslice velocity mismatch can be neglected. Theoretically, the factor $p_R/f_R$ and momentum chirp $dp/d\xi$ can be freely chosen so that the FF propagates at an arbitrary $\beta_f$ which is decoupled from the reference beam velocity. In reality, the accessible values of $\beta_f$ depend on the focusing capability of optics and the maximum momentum chirp permitted by the beamline. The $p_R/f_R m_ec$ of electron beams is typically smaller than $\sim10^4$ m$^{-1}$ for magnetic focusing optics, and can be a few orders of magnitude larger for plasma lenses \cite{Su1990,Pompili2018}.

In the proximity of the focal region, the beam spatial density profile is shaped like a butterfly with the high-density region concentrated only within an effective length $2z_c$ as depicted in Fig. \ref{fig:schematic_ff}. We define $z_c$ as the spacing between the reference slice and a second slice whose cross section area is doubled. To make a large wakefield $z_c\lesssim c/\omega_p$. Assuming a linear chirp $[p(\xi)-p_R]/p_R=h(\xi-\xi_R)$, where $\xi_R$ is the position of the reference slice and $h$ is constant, and neglecting the interslice velocity mismatch, we obtain (see the Supplemental Material)

\begin{equation}
\label{eq:charc_length}
  z_c = |h|^{-1}\sqrt{{\beta^*}^2/(2{\beta^*}^2+f_R^2)},
\end{equation}
where $\beta^*=\sigma_R/\sigma'_R$, $\sigma_R\equiv\sqrt{\langle x^2\rangle}$ and $\sigma'_R\equiv\sqrt{\langle {x'}^2\rangle}$ are the rms size and divergence of the reference slice. A rough criterion for a clear butterfly-shape density profile is $z_c$ should be smaller than the beam length $\sigma_z$. Since $|h|\sim\Delta p/(p_R \sigma_z)$ where $\Delta p/p_R$ is the projected momentum spread, then from Eq. (\ref{eq:charc_length}) and assuming $\Delta p/p_R\ll1$ the criterion $z_c\lesssim\sigma_z$ can be reduced to $\beta^*\lesssim(\Delta p/p_R) f_R$.

The effective diffraction length $L_f$ is the distance within which the FF performs a full intrabeam end-to-end movement. It is straightforward to show that $L_f\simeq f_R \Delta p/p_R$. Typically, we have $z_c\ll\beta^*\lesssim L_f$.

\begin{figure}[htbp]
\includegraphics[width=8.6cm]{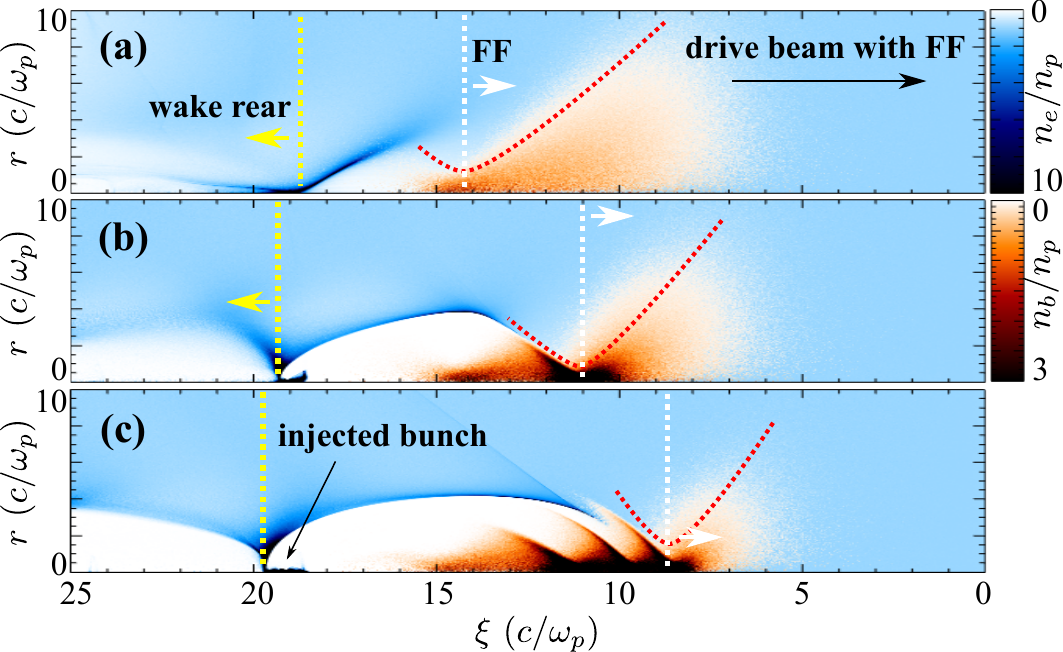}
\caption{\label{fig:wake_evol}
Density distributions of the plasma electrons (blue) and drive beam (red). Snapshots are taken from (a) $\omega_p t=30$ when the drive beam enters the plasma, (b) $\omega_p t=190$, and (c) $\omega_p t=430$. The white and yellow dashed lines mark the positions of FF and the backward expanding ion cavity. The red dashes outline the envelope of the beam density distribution.}
\end{figure}

A high-current electron beam propagating in a plasma can excite a large-amplitude wakefield in the blowout regime when the beam density $n_b$ exceeds the plasma density $n_p$ and the normalized current $\Lambda\equiv4\pi r_e\int n_b r\text{d}r>1$, where $r_e$ is the classical electron radius \cite{Rosenzweig1991,Lu2006a}. The accelerating field $E_z$ is independent of radial position and the transverse focusing force is linear, averting the deterioration of beam slice energy spread and emittance. We demonstrate the proposed injection scheme via particle-in-cell simulations using OSIRIS \cite{Fonseca2002} for a negative-chirped drive beam in the blowout regime, as shown in Fig. \ref{fig:wake_evol}. The reason for not adopting a positive-chirped beam will be discussed later. We have carried out numerous 2D {\it r-z} azimuthally symmetric and a few quasi-3D simulations with two azimuthal modes. We use very fine spatial resolution in both the radial and $z$ (beam propagation direction) directions, $\Delta_r=\Delta_z=0.01c/\omega_p$ to resolve the adiabatic wake expansion. There are 16 macroparticles initialized in each cell to model a uniform plasma with $T_e\sim2$ eV.
We use customized finite-difference solvers \cite{Li2021,Li2017} to eliminate the numerical Cerenkov instability \cite{Xu2013} and spurious space-charge-like fields \cite{Xu2020}. A bi-Gaussian drive beam consisting of $10^6$ macroparticles was initialized at the plane $f_R=10^4 k_p^{-1}$ before the lens with $\gamma_R=1000, h=-0.03, k_p \sigma_z=2$ and $\Lambda=4$, where $k_p\equiv\omega_p/c$ and $\omega_p$ are the plasma wavelength and frequency. At the focal plane ($f_R$ after the lens), the reference slice is focused to $\sigma_R=0.5k_p^{-1}$ with $\sigma_R'=5\times10^{-3}$. The beam particles are tracked using the transfer matrix [Eq. (6) in the Supplemental Material] from the initial plane to where the beam tail is focused, followed by the PIC simulation. With these initial parameters, we know $\beta_f=1.0033$, $k_p z_c=0.67$ and $k_pL_f\sim400$.  

When the drive beam enters the uniform plasma as shown in Fig. \ref{fig:wake_evol}(a), the FF at the beam tail excites a nonlinear plasma wake while the dispersed fraction in front only causes a small perturbation to the plasma. The leading edge of the bubblelike wake follows the superluminal FF as it moves forward. As the focus moves forward an increasing amount of beam charge is contained within the ion channel of the wake, causing an expansion of the blowout radius and wavelength of the ion cavity as shown in Fig. \ref{fig:wake_evol}(b). The receding of the dashed yellow line (moving backward in the speed-of-light frame) dominates over the forward motion of the dashed white line. At this moment the self-injection has been triggered since the injection condition $\beta_\phi<\beta_e$ is satisfied [$\gamma_e\equiv(1-\beta_e^2)^{-1/2}\sim15$ in this case], where $\beta_\phi$ and $\beta_e$ are the wake phase velocity and plasma electron velocity at the rear of the ion cavity. The backward expansion of the wake rear and thus the injection will eventually cease while the FF is still moving forward, as shown in Fig. \ref{fig:wake_evol}(c).

\begin{figure}[htbp]
\centering
\includegraphics[width=0.45\textwidth]{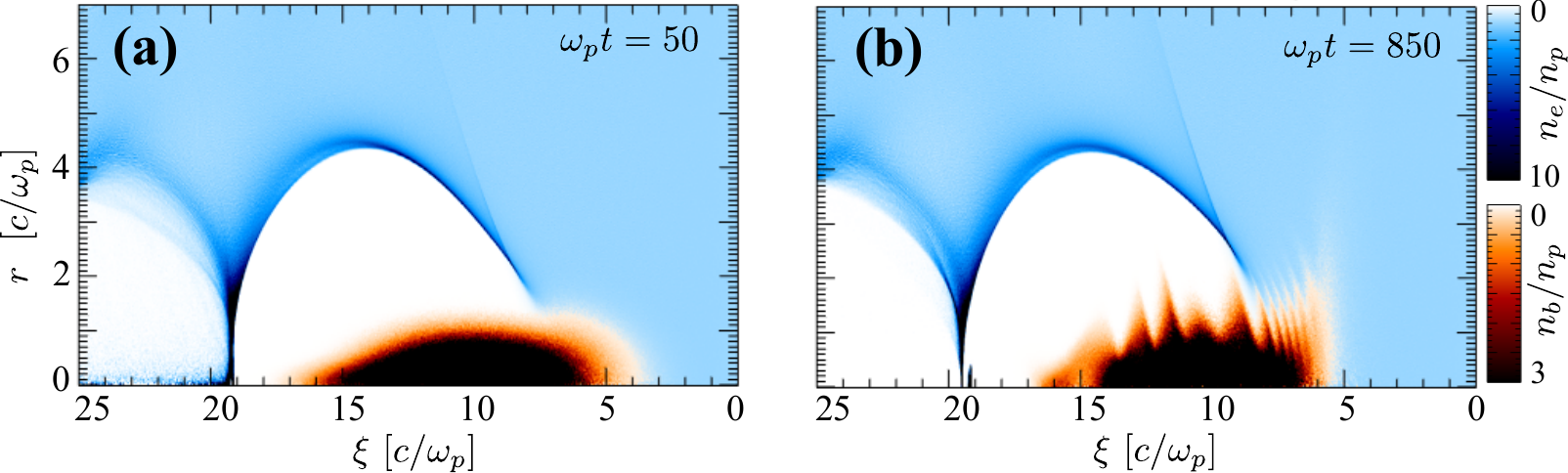}
\caption{Snapshots of simulation with a regular drive beam at (a) $\omega_pt=50$ when just entering the plasma, and (c) $\omega_pt=860$ when the beam energy is nearly exhausted.}
\label{fig:comp_reg_beam}
\end{figure}

As Fig. \ref{fig:wake_evol} makes clear, in order to understand the injection process, we must understand two processes that have opposite effects on $\beta_\phi$: (i) the wake front follows the superluminal FF and travels faster than $c$ (dashed white line), and (ii) as the focus moves forward superluminally more and more beam charge is contained within the wake and confined by the focusing force of the plasma ions. This increases the charge that creates the wake causing a backward expansion of the ion cavity. The latter can dominate for sufficiently large $\Lambda$ such that $\beta_\phi<1$ (dashed yellow line) leading to injection.
For comparison, in Fig. \ref{fig:comp_reg_beam} we also show results from a simulation in which a regular drive beam (without a FF) was used. The drive beam has identical parameters as the example in Fig. \ref{fig:wake_evol} except there is no energy chirp. No significant plasma wake expansion and continuous injection is observed although a very small fraction of background electrons is trapped due to spot size evolution. This indicates that the continuous electron injection is indeed caused by the wake expansion caused by the superluminal FF.

\begin{figure}[htbp]
\includegraphics[width=8.6cm]{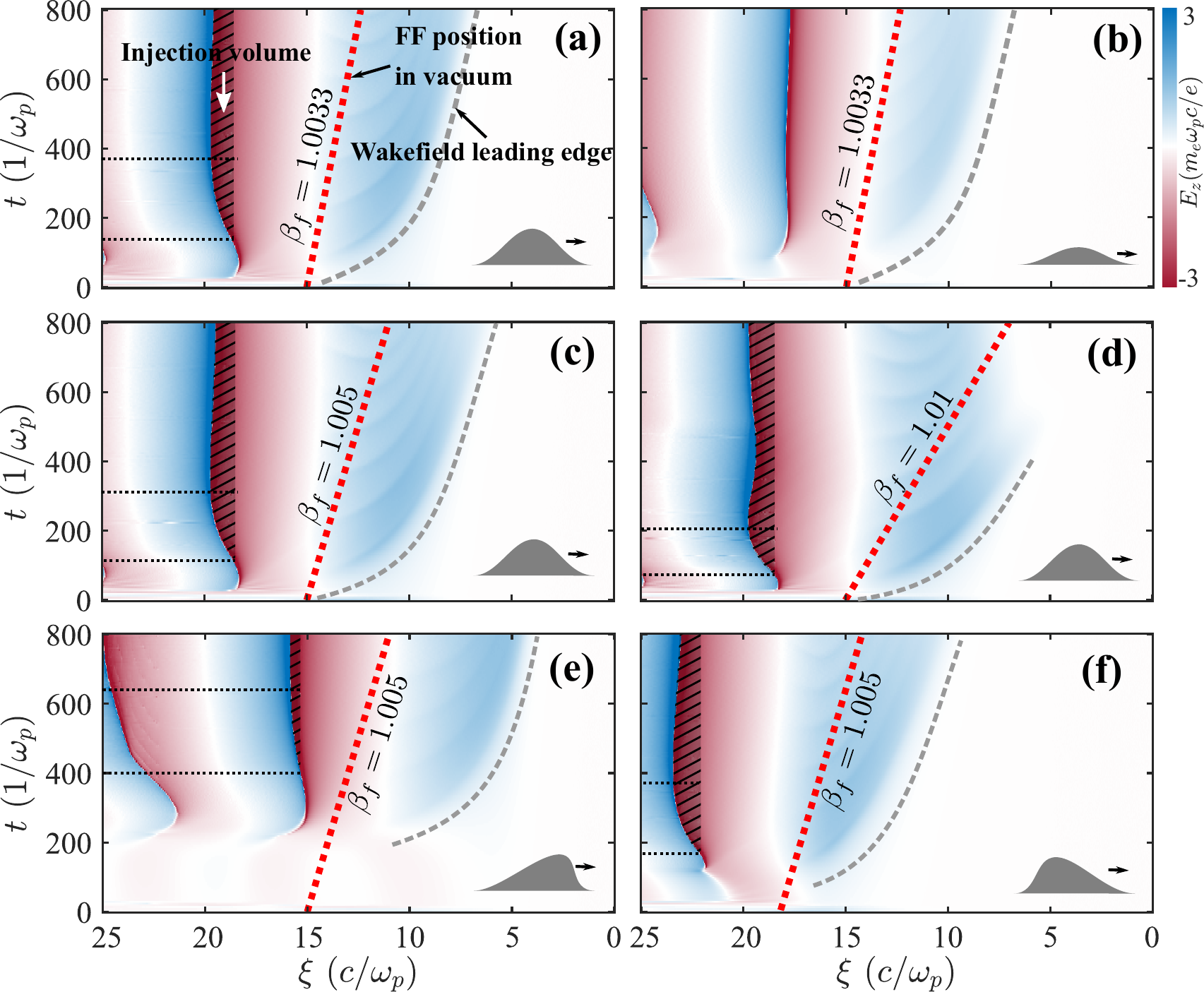}
\caption{\label{fig:ez_evol}
On axis $E_z$ field vs. $\xi$ and $t$. We used $k_p \sigma_R=0.5$, $k_p f_R=10^4$ and $k_p z_c=0.67$ for all the simulations. The results of the (a)--(d) symmetrical beam current profiles with $k_p \sigma_z=2$ and various $\beta_f$ [$\Lambda=4$ for (a)(c)(d) and $\Lambda=2$ for (b)], (e) forward-tilt beam current profile with $k_p \sigma_{z1}=3$, $k_p \sigma_{z2}=1$, $\Lambda=4$ and $\beta_f=1.005$, and (f) backward-tilt beam current profile with $k_p \sigma_{z1}=1$, $k_p \sigma_{z2}=3$, $\Lambda=4$ and $\beta_f=1.005$.}
\end{figure}

The physics of self-injection and the competing processes are further revealed in Fig. \ref{fig:ez_evol}. Each frame corresponds to different drive beam parameters and for each case we show the evolution of the on axis $E_z$ wakefield, the leading edge of the wake (gray dashed line), FF location in vacuum (red dashed line) and the region where the injected beam resides (shaded). In the speed-of-light coordinate ($\xi$) a vertical line corresponds to a point moving at $c$ while a line with a positive (negative) slope corresponds to superluminal (subluminal) speed. Injection occurs at the back of the first accelerating bucket of the wake where the field changes from accelerating (red) to decelerating (blue). Injection can occur when a negative slope ($\beta_\phi<1$) develops at the rear of the wake and stops when the slope becomes almost vertical ($\beta_\phi=1$). In each frame it is clear that while the front of the wake moves superluminally due to the moving focus the electron injection occurs at the back of the wake that moves subluminally. In Fig. \ref{fig:ez_evol}(a) the obvious backward wake expansion only occurs when $t\lesssim380\omega_p^{-1}$ and thereafter the injection volume remains almost unchanged, indicating a stable subsequent acceleration with little beam loss. In Fig. \ref{fig:ez_evol}(b) where $\Lambda=2$, the drive beam does not contain a sufficient charge to expand the wake to reduce $\beta_\phi$. In all cases, the leading edge of $E_z$ does not coincide with the FF trajectory in vacuum. It first speeds up due to the focusing from the small-amplitude wake created by the low-density part of beam in front of the FF, and then moves at a speed $\sim\beta_f$ as the amount of charge before the FF diminishes and the wake-induced focusing weakens.

The duration and thereby the charge and energy spread of the injected beam can be controlled by tuning $\beta_f$ and shaping the current profile of the drive beams. Figures \ref{fig:ez_evol}(c) and \ref{fig:ez_evol}(d) show results for $\beta_f=1.005$ and $\beta_f=1.01$ while $\sigma_R,~z_c,~f_R$, and $\Lambda$ were kept the same as in Fig. \ref{fig:ez_evol}(a). To control $\beta_f$ we changed the linear chirp coefficient $h$. As $\beta_f$ is increased, the injection duration (spacing between the black dashed lines) shortens, which leads to a different ``optimal'' energy of the accelerated bunches (this will be discussed later). Figure \ref{fig:ez_evol}(e)--\ref{fig:ez_evol}(f) show the injection processes using forward-tilted and backward-tilted asymmetric Gaussian current profiles $I(\xi)=(1/2) I_A \Lambda\{\Theta(\xi-\xi_R)\exp[-(\xi-\xi_R)^2/(2\sigma_{z1}^2)]+\Theta(\xi_R-\xi)\exp[-(\xi-\xi_R)^2/(2\sigma_{z2}^2)]\}$ where $\Theta(x)$ is the Heaviside step function and $I_A=17$ kA is the Alfv\'en current. For the forward-tilted case, the current at the back of the beam which is initially focused is relatively low so that the bubble expansion is delayed, and the injection occurs later as expected.

\begin{figure}[htbp]
\includegraphics[width=8.6cm]{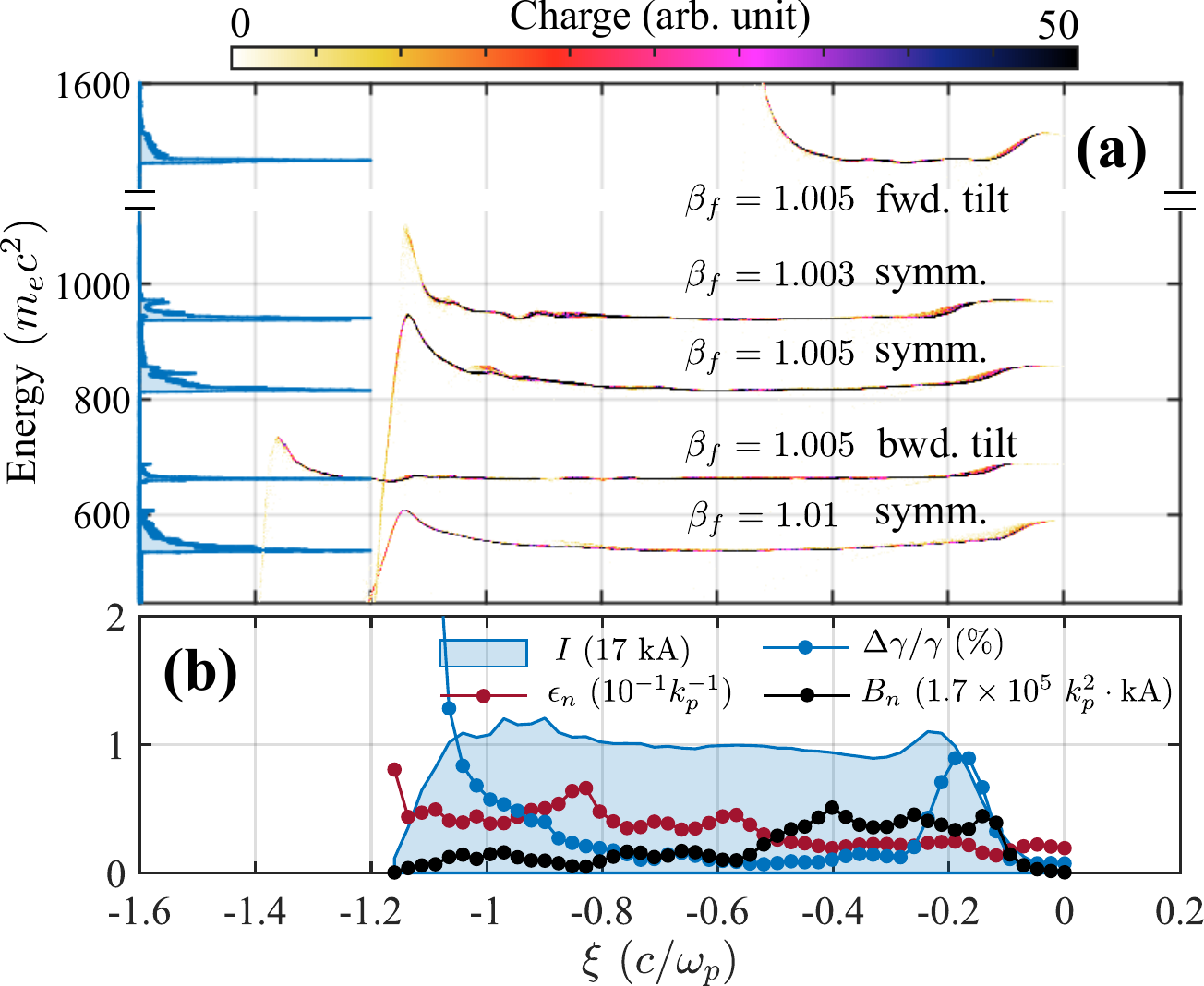}
\caption{\label{fig:beam_quality}
  The longitudinal phase spaces and the energy spectrum of each accelerated beam following injection induced by superluminal FF examples shown in Fig. \ref{fig:ez_evol}. The energy spectra on the lhs have been normalized for a better visualization. (b) The sliced beam properties of the injected beam in Fig. \ref{fig:ez_evol}(a) or Fig. \ref{fig:wake_evol}.
}
\end{figure}

The optimal beam energy $\gamma_\text{opt}$ is related to the backward expansion rate of the ion cavity and is thus tunable through $\beta_f$ and current shaping of the drive beam. Here, the optimal energy means the average beam energy when the projected energy spread reaches the minimum \cite{Xu2017}. During the injection process, the electrons injected earlier have a higher energy than that of the electrons injected at the end. An initial energy chirp $\Delta\gamma m_e c^2\sim e\bar{E}_z L_\text{inj}$ emerges immediately after the injection where $L_\text{inj}$ is the distance over which the injection occurs and $\bar{E}_z$ is the average accelerating field felt by the beam. Due to the shape of the blowout regime, the accelerating gradient experienced by the beam tail is larger than the head; hence, the chirp will be eliminated after an optimum acceleration distance $L_\text{opt}\sim\bar{E}_z/(\Delta E_z ) L_\text{inj}$ where $\Delta E_z$ is the difference of the accelerating field amplitude felt by both ends of the beam. Without beam loading \cite{Tzoufras2008,Dalichaouch2021} $\Delta E_z=m_e\omega_p^2\sigma_z/(2e)$ and this can be used as a lower bound. The optimal energy can be estimated as the sum of the energy gain during and after the injection, i.e., $\gamma_\text{opt} m_e c^2\sim e\bar{E}_z L_\text{inj} (1/2+\bar{E}_z/\Delta E_z)$. For short injected bunches $\bar{E}_z/\Delta E_z\sim R/\sigma_z\gg 1/2$ where $R$ is the blowout radius of the wake. Since the ion cavity expands at a rate $w_\phi=1-\beta_\phi\sim \sigma_z/L_\text{inj}$, we know that $\gamma_\text{opt}\propto \bar{E}_z R/w_\phi$. Figure \ref{fig:beam_quality}(a) shows the $p_z$-$\xi$ phase space of the injected bunches in all the cases of Fig. \ref{fig:ez_evol}. The projected energy spreads are $<1\%$ in all cases. For the central (middle half) portion of the beam where the energy curves upward in the front and back parts are excluded, the energy spread can be as low as $\sim0.3\%$. 

Comparing the three cases with symmetric current profiles, we see that a larger $\beta_f$ leads to a larger $w_\phi$ and hence lower $\gamma_\text{opt}$, which is consistent with the previous analysis. This reasoning can also be applied to determine how to tune $\gamma_\text{opt}$ for the tilted current profiles. The rear of the wake expands faster in the backward-tilted case because more charge is trapped sooner. As the charge in the front of then beam is trapped the expansion rate can drop below the threshold for trapping. As a result, this also leads to shorter $L_\text{inj}$ even though $\beta_f$ is identical. Thus, the $\gamma_\text{opt}$ for the backward-tilted current profile is lower.

The simulation results show that the proposed injection scheme can generate a high-quality electron bunch which simultaneously possesses an ultralow normalized emittance and energy spread, and high current and thus very high normalized brightness.
Figure \ref{fig:beam_quality}(b) shows the slice energy spread, current, emittance and brightness of the injected beam as a function of $\xi$, which is taken from the simulation in Fig. \ref{fig:ez_evol}(a). Similar results can be obtained for other simulations. In this example, the self-injected beam has an average current of 17 kA, a slice normalized emittance of $\epsilon_n\sim0.02~k_p^{-1}$, a slice energy spread $\sim0.1\%$, and a peak normalized brightness of $B_n[\text{\Bunit}]\sim8.5\times10^7 k_p^2[\text{m$^{-1}$}]$.

Each simulation corresponds to a family of beam and plasma parameters with the same normalized values. The values of $\epsilon_n$ and $B_n$ scale as $n_p^{-1/2}$ and $n_p$ respectively \cite{Xu2017}, for example, $\epsilon_n=34$ nm and $B_n=3\times10^{19}$ \Bunit~for $n_p=1\times10^{19}$ cm$^{-3}$. Current state-of-the-art conventional accelerators are anticipated to produce 50$\sim$150 kA, $\sim$3 fs electron bunches \cite{Yakimenko2019}.
Such ultrashort bunch length allows for the operation of a PWFA for $n_p\sim10^{20}$ cm$^{-3}$.
This indicates that generating self-injected beams with $B_n\sim10^{20}$ \Bunit~using the proposed injection scheme may be possible. In the Supplemental Material, we provide estimates of FEL output using these ultrabright beams according to 1D FEL theory \cite{Freund1992}. It indicates that improvements to the brightness of FEL radiation are achievable with a much shorter undulator. 

As discussed earlier the continued focusing of the drive beam as the FF propagates forward is critical. It leads to a continued increase in beam charge contained by the wake and the resulting expansion of the wake. As shown in the Supplemental material, a superluminal FF enables such focusing whereas the subluminal focusing does not. In the subluminal case, as the focus moves backward with respect to $c$, the beam head diffracts and less and less charge resides in the ion channel causing its length to shorten and the rear of the wake to move forward. The wake is eventually terminated.
We found that the pump depletion length, $L_\text{pd}$, of drive beams not only depends nearly linearly on the initial energy but also on $\beta_f$. With an initial FF position $2.5\sigma_z$ behind the reference beam slice, the observed $L_\text{pd}$ are $970k_p^{-1}$, $995k_p^{-1}$ and $1030k_p^{-1}$ for $\beta_f=1.0033,~1.005$ and 1.01, respectively. A longer $L_\text{pd}$ apparently leads to a larger final energy gain, but the gain at the optimal acceleration distance where the energy chirp is minimized is primarily determined by $\beta_f$ according to the simulation results. A general quantitative analysis for $L_\text{pd}$ and energy gain is subtle for FF beams and we leave this for future work.

In conclusion, we have shown that ultrabright electron bunches can be generated by a superluminal FF in a uniform plasma by sending a negatively chirped charged particle beam through a focusing optic. This method is feasible with the current state-of-the-art electron accelerators and only requires a relatively simple experimental configuration. 

\begin{acknowledgments}
This work was supported by the US Department of Energy through a SciDAC FNAL Subcontract No. 644405 and Grant No. DE-SC0010064, and the US National Science Foundation Grants No. 1806046 and No. 2003354. The simulations were performed on the UCLA Hoffman 2 and Dawson 2 Clusters, and the computing resources of the National Energy Research Scientific Computing Center. 
\end{acknowledgments}

\bibliography{ref}

\begin{thebibliography}{43}%
\makeatletter
\providecommand \@ifxundefined [1]{%
 \@ifx{#1\undefined}
}%
\providecommand \@ifnum [1]{%
 \ifnum #1\expandafter \@firstoftwo
 \else \expandafter \@secondoftwo
 \fi
}%
\providecommand \@ifx [1]{%
 \ifx #1\expandafter \@firstoftwo
 \else \expandafter \@secondoftwo
 \fi
}%
\providecommand \natexlab [1]{#1}%
\providecommand \enquote  [1]{``#1''}%
\providecommand \bibnamefont  [1]{#1}%
\providecommand \bibfnamefont [1]{#1}%
\providecommand \citenamefont [1]{#1}%
\providecommand \href@noop [0]{\@secondoftwo}%
\providecommand \href [0]{\begingroup \@sanitize@url \@href}%
\providecommand \@href[1]{\@@startlink{#1}\@@href}%
\providecommand \@@href[1]{\endgroup#1\@@endlink}%
\providecommand \@sanitize@url [0]{\catcode `\\12\catcode `\$12\catcode
  `\&12\catcode `\#12\catcode `\^12\catcode `\_12\catcode `\%12\relax}%
\providecommand \@@startlink[1]{}%
\providecommand \@@endlink[0]{}%
\providecommand \url  [0]{\begingroup\@sanitize@url \@url }%
\providecommand \@url [1]{\endgroup\@href {#1}{\urlprefix }}%
\providecommand \urlprefix  [0]{URL }%
\providecommand \Eprint [0]{\href }%
\providecommand \doibase [0]{https://doi.org/}%
\providecommand \selectlanguage [0]{\@gobble}%
\providecommand \bibinfo  [0]{\@secondoftwo}%
\providecommand \bibfield  [0]{\@secondoftwo}%
\providecommand \translation [1]{[#1]}%
\providecommand \BibitemOpen [0]{}%
\providecommand \bibitemStop [0]{}%
\providecommand \bibitemNoStop [0]{.\EOS\space}%
\providecommand \EOS [0]{\spacefactor3000\relax}%
\providecommand \BibitemShut  [1]{\csname bibitem#1\endcsname}%
\let\auto@bib@innerbib\@empty
%</preamble>
\bibitem [{\citenamefont {Joshi}\ and\ \citenamefont
  {Katsouleas}(2003)}]{Joshi2003}%
  \BibitemOpen
  \bibfield  {author} {\bibinfo {author} {\bibfnamefont {C.}~\bibnamefont
  {Joshi}}\ and\ \bibinfo {author} {\bibfnamefont {T.}~\bibnamefont
  {Katsouleas}},\ }\bibfield  {title} {\bibinfo {title} {{Plasma accelerators
  at the energy frontier and on tabletops}},\ }\href@noop {} {\bibfield
  {journal} {\bibinfo  {journal} {Physics Today}\ }\textbf {\bibinfo {volume}
  {56}},\ \bibinfo {pages} {47} (\bibinfo {year} {2003})}\BibitemShut {NoStop}%
\bibitem [{\citenamefont {Joshi}\ \emph {et~al.}(2020)\citenamefont {Joshi},
  \citenamefont {Corde},\ and\ \citenamefont {Mori}}]{Joshi2020}%
  \BibitemOpen
  \bibfield  {author} {\bibinfo {author} {\bibfnamefont {C.}~\bibnamefont
  {Joshi}}, \bibinfo {author} {\bibfnamefont {S.}~\bibnamefont {Corde}},\ and\
  \bibinfo {author} {\bibfnamefont {W.~B.}\ \bibnamefont {Mori}},\ }\bibfield
  {title} {\bibinfo {title} {{Perspectives on the generation of electron beams
  from plasma-based accelerators and their near and long term applications}},\
  }\href@noop {} {\bibfield  {journal} {\bibinfo  {journal} {Physics of
  Plasmas}\ }\textbf {\bibinfo {volume} {27}},\ \bibinfo {pages} {070602}
  (\bibinfo {year} {2020})}\BibitemShut {NoStop}%
\bibitem [{\citenamefont {Wang}\ \emph {et~al.}(2021)\citenamefont {Wang},
  \citenamefont {Feng}, \citenamefont {Ke}, \citenamefont {Yu}, \citenamefont
  {Xu}, \citenamefont {Qi}, \citenamefont {Chen}, \citenamefont {Qin},
  \citenamefont {Zhang}, \citenamefont {Fang}, \citenamefont {Liu},
  \citenamefont {Jiang}, \citenamefont {Wang}, \citenamefont {Wang},
  \citenamefont {Yang}, \citenamefont {Wu}, \citenamefont {Leng}, \citenamefont
  {Liu}, \citenamefont {Li},\ and\ \citenamefont {Xu}}]{Wang2021}%
  \BibitemOpen
  \bibfield  {author} {\bibinfo {author} {\bibfnamefont {W.}~\bibnamefont
  {Wang}}, \bibinfo {author} {\bibfnamefont {K.}~\bibnamefont {Feng}}, \bibinfo
  {author} {\bibfnamefont {L.}~\bibnamefont {Ke}}, \bibinfo {author}
  {\bibfnamefont {C.}~\bibnamefont {Yu}}, \bibinfo {author} {\bibfnamefont
  {Y.}~\bibnamefont {Xu}}, \bibinfo {author} {\bibfnamefont {R.}~\bibnamefont
  {Qi}}, \bibinfo {author} {\bibfnamefont {Y.}~\bibnamefont {Chen}}, \bibinfo
  {author} {\bibfnamefont {Z.}~\bibnamefont {Qin}}, \bibinfo {author}
  {\bibfnamefont {Z.}~\bibnamefont {Zhang}}, \bibinfo {author} {\bibfnamefont
  {M.}~\bibnamefont {Fang}}, \bibinfo {author} {\bibfnamefont {J.}~\bibnamefont
  {Liu}}, \bibinfo {author} {\bibfnamefont {K.}~\bibnamefont {Jiang}}, \bibinfo
  {author} {\bibfnamefont {H.}~\bibnamefont {Wang}}, \bibinfo {author}
  {\bibfnamefont {C.}~\bibnamefont {Wang}}, \bibinfo {author} {\bibfnamefont
  {X.}~\bibnamefont {Yang}}, \bibinfo {author} {\bibfnamefont {F.}~\bibnamefont
  {Wu}}, \bibinfo {author} {\bibfnamefont {Y.}~\bibnamefont {Leng}}, \bibinfo
  {author} {\bibfnamefont {J.}~\bibnamefont {Liu}}, \bibinfo {author}
  {\bibfnamefont {R.}~\bibnamefont {Li}},\ and\ \bibinfo {author}
  {\bibfnamefont {Z.}~\bibnamefont {Xu}},\ }\bibfield  {title} {\bibinfo
  {title} {{Free-electron lasing at 27 nanometres based on a laser wakefield
  accelerator}},\ }\href@noop {} {\bibfield  {journal} {\bibinfo  {journal}
  {Nature}\ }\textbf {\bibinfo {volume} {595}},\ \bibinfo {pages} {516}
  (\bibinfo {year} {2021})}\BibitemShut {NoStop}%
\bibitem [{\citenamefont {Xu}\ \emph {et~al.}(2020{\natexlab{a}})\citenamefont
  {Xu}, \citenamefont {Li}, \citenamefont {Tsung}, \citenamefont {Miller},
  \citenamefont {Yakimenko}, \citenamefont {Hogan}, \citenamefont {Joshi},\
  and\ \citenamefont {Mori}}]{Xu2021}%
  \BibitemOpen
  \bibfield  {author} {\bibinfo {author} {\bibfnamefont {X.}~\bibnamefont
  {Xu}}, \bibinfo {author} {\bibfnamefont {F.}~\bibnamefont {Li}}, \bibinfo
  {author} {\bibfnamefont {F.~S.}\ \bibnamefont {Tsung}}, \bibinfo {author}
  {\bibfnamefont {K.}~\bibnamefont {Miller}}, \bibinfo {author} {\bibfnamefont
  {V.}~\bibnamefont {Yakimenko}}, \bibinfo {author} {\bibfnamefont {M.~J.}\
  \bibnamefont {Hogan}}, \bibinfo {author} {\bibfnamefont {C.}~\bibnamefont
  {Joshi}},\ and\ \bibinfo {author} {\bibfnamefont {W.~B.}\ \bibnamefont
  {Mori}},\ }\href@noop {} {\bibinfo {title} {{Generation and acceleration of
  high brightness electrons beams bunched at X-ray wavelengths using
  plasma-based acceleration}}} (\bibinfo {year} {2020}{\natexlab{a}}),\ \Eprint
  {https://arxiv.org/abs/2010.16081} {arXiv:2010.16081 [physics.acc-ph]}
  \BibitemShut {NoStop}%
\bibitem [{\citenamefont {Pellegrini}\ \emph {et~al.}(2016)\citenamefont
  {Pellegrini}, \citenamefont {Marinelli},\ and\ \citenamefont
  {Reiche}}]{Pellegrini2016}%
  \BibitemOpen
  \bibfield  {author} {\bibinfo {author} {\bibfnamefont {C.}~\bibnamefont
  {Pellegrini}}, \bibinfo {author} {\bibfnamefont {A.}~\bibnamefont
  {Marinelli}},\ and\ \bibinfo {author} {\bibfnamefont {S.}~\bibnamefont
  {Reiche}},\ }\bibfield  {title} {\bibinfo {title} {{The physics of x-ray
  free-electron lasers}},\ }\href
  {http://link.aps.org/doi/10.1103/RevModPhys.88.015006} {\bibfield  {journal}
  {\bibinfo  {journal} {Reviews of Modern Physics}\ }\textbf {\bibinfo {volume}
  {88}},\ \bibinfo {pages} {015006} (\bibinfo {year} {2016})}\BibitemShut
  {NoStop}%
\bibitem [{\citenamefont {Chen}\ \emph {et~al.}(2006)\citenamefont {Chen},
  \citenamefont {Sheng}, \citenamefont {Ma},\ and\ \citenamefont
  {Zhang}}]{Chen2006}%
  \BibitemOpen
  \bibfield  {author} {\bibinfo {author} {\bibfnamefont {M.}~\bibnamefont
  {Chen}}, \bibinfo {author} {\bibfnamefont {Z.~M.}\ \bibnamefont {Sheng}},
  \bibinfo {author} {\bibfnamefont {Y.~Y.}\ \bibnamefont {Ma}},\ and\ \bibinfo
  {author} {\bibfnamefont {J.}~\bibnamefont {Zhang}},\ }\bibfield  {title}
  {\bibinfo {title} {{Electron injection and trapping in a laser wakefield by
  field ionization to high-charge states of gases}},\ }\href@noop {} {\bibfield
   {journal} {\bibinfo  {journal} {Journal of Applied Physics}\ }\textbf
  {\bibinfo {volume} {99}},\ \bibinfo {pages} {056109} (\bibinfo {year}
  {2006})}\BibitemShut {NoStop}%
\bibitem [{\citenamefont {Oz}\ \emph {et~al.}(2007)\citenamefont {Oz},
  \citenamefont {Deng}, \citenamefont {Katsouleas}, \citenamefont {Muggli},
  \citenamefont {Barnes}, \citenamefont {Blumenfeld}, \citenamefont {Decker},
  \citenamefont {Emma}, \citenamefont {Hogan}, \citenamefont {Ischebeck},
  \citenamefont {Iverson}, \citenamefont {Kirby}, \citenamefont {Krejcik},
  \citenamefont {O'connell}, \citenamefont {Siemann}, \citenamefont {Walz},
  \citenamefont {Auerbach}, \citenamefont {Clayton}, \citenamefont {Huang},
  \citenamefont {Johnson}, \citenamefont {Joshi}, \citenamefont {Lu},
  \citenamefont {Marsh}, \citenamefont {Mori},\ and\ \citenamefont
  {Zhou}}]{Oz2007}%
  \BibitemOpen
  \bibfield  {author} {\bibinfo {author} {\bibfnamefont {E.}~\bibnamefont
  {Oz}}, \bibinfo {author} {\bibfnamefont {S.}~\bibnamefont {Deng}}, \bibinfo
  {author} {\bibfnamefont {T.}~\bibnamefont {Katsouleas}}, \bibinfo {author}
  {\bibfnamefont {P.}~\bibnamefont {Muggli}}, \bibinfo {author} {\bibfnamefont
  {C.~D.}\ \bibnamefont {Barnes}}, \bibinfo {author} {\bibfnamefont
  {I.}~\bibnamefont {Blumenfeld}}, \bibinfo {author} {\bibfnamefont {F.~J.}\
  \bibnamefont {Decker}}, \bibinfo {author} {\bibfnamefont {P.}~\bibnamefont
  {Emma}}, \bibinfo {author} {\bibfnamefont {M.~J.}\ \bibnamefont {Hogan}},
  \bibinfo {author} {\bibfnamefont {R.}~\bibnamefont {Ischebeck}}, \bibinfo
  {author} {\bibfnamefont {R.~H.}\ \bibnamefont {Iverson}}, \bibinfo {author}
  {\bibfnamefont {N.}~\bibnamefont {Kirby}}, \bibinfo {author} {\bibfnamefont
  {P.}~\bibnamefont {Krejcik}}, \bibinfo {author} {\bibfnamefont
  {C.}~\bibnamefont {O'connell}}, \bibinfo {author} {\bibfnamefont {R.~H.}\
  \bibnamefont {Siemann}}, \bibinfo {author} {\bibfnamefont {D.}~\bibnamefont
  {Walz}}, \bibinfo {author} {\bibfnamefont {D.}~\bibnamefont {Auerbach}},
  \bibinfo {author} {\bibfnamefont {C.~E.}\ \bibnamefont {Clayton}}, \bibinfo
  {author} {\bibfnamefont {C.}~\bibnamefont {Huang}}, \bibinfo {author}
  {\bibfnamefont {D.~K.}\ \bibnamefont {Johnson}}, \bibinfo {author}
  {\bibfnamefont {C.}~\bibnamefont {Joshi}}, \bibinfo {author} {\bibfnamefont
  {W.}~\bibnamefont {Lu}}, \bibinfo {author} {\bibfnamefont {K.~A.}\
  \bibnamefont {Marsh}}, \bibinfo {author} {\bibfnamefont {W.~B.}\ \bibnamefont
  {Mori}},\ and\ \bibinfo {author} {\bibfnamefont {M.}~\bibnamefont {Zhou}},\
  }\bibfield  {title} {\bibinfo {title} {{Ionization-induced electron trapping
  in ultrarelativistic plasma wakes}},\ }\href@noop {} {\bibfield  {journal}
  {\bibinfo  {journal} {Physical Review Letters}\ }\textbf {\bibinfo {volume}
  {98}},\ \bibinfo {pages} {084801} (\bibinfo {year} {2007})}\BibitemShut
  {NoStop}%
\bibitem [{\citenamefont {Pak}\ \emph {et~al.}(2010)\citenamefont {Pak},
  \citenamefont {Marsh}, \citenamefont {Martins}, \citenamefont {Lu},
  \citenamefont {Mori},\ and\ \citenamefont {Joshi}}]{Pak2010}%
  \BibitemOpen
  \bibfield  {author} {\bibinfo {author} {\bibfnamefont {A.}~\bibnamefont
  {Pak}}, \bibinfo {author} {\bibfnamefont {K.~A.}\ \bibnamefont {Marsh}},
  \bibinfo {author} {\bibfnamefont {S.~F.}\ \bibnamefont {Martins}}, \bibinfo
  {author} {\bibfnamefont {W.}~\bibnamefont {Lu}}, \bibinfo {author}
  {\bibfnamefont {W.~B.}\ \bibnamefont {Mori}},\ and\ \bibinfo {author}
  {\bibfnamefont {C.}~\bibnamefont {Joshi}},\ }\bibfield  {title} {\bibinfo
  {title} {{Injection and trapping of tunnel-ionized electrons into
  laser-produced wakes}},\ }\href@noop {} {\bibfield  {journal} {\bibinfo
  {journal} {Physical Review Letters}\ }\textbf {\bibinfo {volume} {104}},\
  \bibinfo {pages} {025003} (\bibinfo {year} {2010})}\BibitemShut {NoStop}%
\bibitem [{\citenamefont {Hidding}\ \emph {et~al.}(2012)\citenamefont
  {Hidding}, \citenamefont {Pretzler}, \citenamefont {Rosenzweig},
  \citenamefont {K{\"{o}}nigstein}, \citenamefont {Schiller},\ and\
  \citenamefont {Bruhwiler}}]{Hidding2012}%
  \BibitemOpen
  \bibfield  {author} {\bibinfo {author} {\bibfnamefont {B.}~\bibnamefont
  {Hidding}}, \bibinfo {author} {\bibfnamefont {G.}~\bibnamefont {Pretzler}},
  \bibinfo {author} {\bibfnamefont {J.~B.}\ \bibnamefont {Rosenzweig}},
  \bibinfo {author} {\bibfnamefont {T.}~\bibnamefont {K{\"{o}}nigstein}},
  \bibinfo {author} {\bibfnamefont {D.}~\bibnamefont {Schiller}},\ and\
  \bibinfo {author} {\bibfnamefont {D.~L.}\ \bibnamefont {Bruhwiler}},\
  }\bibfield  {title} {\bibinfo {title} {{Ultracold electron bunch generation
  via plasma photocathode emission and acceleration in a beam-driven plasma
  blowout}},\ }\href@noop {} {\bibfield  {journal} {\bibinfo  {journal}
  {Physical Review Letters}\ }\textbf {\bibinfo {volume} {108}},\ \bibinfo
  {pages} {035001} (\bibinfo {year} {2012})}\BibitemShut {NoStop}%
\bibitem [{\citenamefont {Li}\ \emph {et~al.}(2013)\citenamefont {Li},
  \citenamefont {Hua}, \citenamefont {Xu}, \citenamefont {Zhang}, \citenamefont
  {Yan}, \citenamefont {Du}, \citenamefont {Huang}, \citenamefont {Chen},
  \citenamefont {Tang}, \citenamefont {Lu}, \citenamefont {Joshi},
  \citenamefont {Mori},\ and\ \citenamefont {Gu}}]{Li2013}%
  \BibitemOpen
  \bibfield  {author} {\bibinfo {author} {\bibfnamefont {F.}~\bibnamefont
  {Li}}, \bibinfo {author} {\bibfnamefont {J.~F.}\ \bibnamefont {Hua}},
  \bibinfo {author} {\bibfnamefont {X.~L.}\ \bibnamefont {Xu}}, \bibinfo
  {author} {\bibfnamefont {C.~J.}\ \bibnamefont {Zhang}}, \bibinfo {author}
  {\bibfnamefont {L.~X.}\ \bibnamefont {Yan}}, \bibinfo {author} {\bibfnamefont
  {Y.~C.}\ \bibnamefont {Du}}, \bibinfo {author} {\bibfnamefont {W.~H.}\
  \bibnamefont {Huang}}, \bibinfo {author} {\bibfnamefont {H.~B.}\ \bibnamefont
  {Chen}}, \bibinfo {author} {\bibfnamefont {C.~X.}\ \bibnamefont {Tang}},
  \bibinfo {author} {\bibfnamefont {W.}~\bibnamefont {Lu}}, \bibinfo {author}
  {\bibfnamefont {C.}~\bibnamefont {Joshi}}, \bibinfo {author} {\bibfnamefont
  {W.~B.}\ \bibnamefont {Mori}},\ and\ \bibinfo {author} {\bibfnamefont
  {Y.~Q.}\ \bibnamefont {Gu}},\ }\bibfield  {title} {\bibinfo {title}
  {{Generating high-brightness electron beams via ionization injection by
  transverse colliding lasers in a plasma-Wakefield accelerator}},\ }\href@noop
  {} {\bibfield  {journal} {\bibinfo  {journal} {Physical Review Letters}\
  }\textbf {\bibinfo {volume} {111}},\ \bibinfo {pages} {015003} (\bibinfo
  {year} {2013})}\BibitemShut {NoStop}%
\bibitem [{\citenamefont {Martinez De La~Ossa}\ \emph
  {et~al.}(2013)\citenamefont {Martinez De La~Ossa}, \citenamefont {Grebenyuk},
  \citenamefont {Mehrling}, \citenamefont {Schaper},\ and\ \citenamefont
  {Osterhoff}}]{MartinezDeLaOssa2013}%
  \BibitemOpen
  \bibfield  {author} {\bibinfo {author} {\bibfnamefont {A.}~\bibnamefont
  {Martinez De La~Ossa}}, \bibinfo {author} {\bibfnamefont {J.}~\bibnamefont
  {Grebenyuk}}, \bibinfo {author} {\bibfnamefont {T.}~\bibnamefont {Mehrling}},
  \bibinfo {author} {\bibfnamefont {L.}~\bibnamefont {Schaper}},\ and\ \bibinfo
  {author} {\bibfnamefont {J.}~\bibnamefont {Osterhoff}},\ }\bibfield  {title}
  {\bibinfo {title} {{High-quality electron beams from beam-driven plasma
  accelerators by wakefield-induced ionization injection}},\ }\href@noop {}
  {\bibfield  {journal} {\bibinfo  {journal} {Physical Review Letters}\
  }\textbf {\bibinfo {volume} {111}},\ \bibinfo {pages} {245003} (\bibinfo
  {year} {2013})}\BibitemShut {NoStop}%
\bibitem [{\citenamefont {Xu}\ \emph {et~al.}(2014)\citenamefont {Xu},
  \citenamefont {Wu}, \citenamefont {Zhang}, \citenamefont {Li}, \citenamefont
  {Wan}, \citenamefont {Hua}, \citenamefont {Pai}, \citenamefont {Lu},
  \citenamefont {Yu}, \citenamefont {Joshi},\ and\ \citenamefont
  {Mori}}]{Xu2014}%
  \BibitemOpen
  \bibfield  {author} {\bibinfo {author} {\bibfnamefont {X.~L.}\ \bibnamefont
  {Xu}}, \bibinfo {author} {\bibfnamefont {Y.~P.}\ \bibnamefont {Wu}}, \bibinfo
  {author} {\bibfnamefont {C.~J.}\ \bibnamefont {Zhang}}, \bibinfo {author}
  {\bibfnamefont {F.}~\bibnamefont {Li}}, \bibinfo {author} {\bibfnamefont
  {Y.}~\bibnamefont {Wan}}, \bibinfo {author} {\bibfnamefont {J.~F.}\
  \bibnamefont {Hua}}, \bibinfo {author} {\bibfnamefont {C.~H.}\ \bibnamefont
  {Pai}}, \bibinfo {author} {\bibfnamefont {W.}~\bibnamefont {Lu}}, \bibinfo
  {author} {\bibfnamefont {P.}~\bibnamefont {Yu}}, \bibinfo {author}
  {\bibfnamefont {C.}~\bibnamefont {Joshi}},\ and\ \bibinfo {author}
  {\bibfnamefont {W.~B.}\ \bibnamefont {Mori}},\ }\bibfield  {title} {\bibinfo
  {title} {{Low emittance electron beam generation from a laser wakefield
  accelerator using two laser pulses with different wavelengths}},\ }\href@noop
  {} {\bibfield  {journal} {\bibinfo  {journal} {Physical Review Special Topics
  - Accelerators and Beams}\ }\textbf {\bibinfo {volume} {17}},\ \bibinfo
  {pages} {061301} (\bibinfo {year} {2014})}\BibitemShut {NoStop}%
\bibitem [{\citenamefont {Bulanov}\ \emph {et~al.}(1998)\citenamefont
  {Bulanov}, \citenamefont {Naumova}, \citenamefont {Pegoraro},\ and\
  \citenamefont {Sakai}}]{Bulanov1998}%
  \BibitemOpen
  \bibfield  {author} {\bibinfo {author} {\bibfnamefont {S.}~\bibnamefont
  {Bulanov}}, \bibinfo {author} {\bibfnamefont {N.}~\bibnamefont {Naumova}},
  \bibinfo {author} {\bibfnamefont {F.}~\bibnamefont {Pegoraro}},\ and\
  \bibinfo {author} {\bibfnamefont {J.}~\bibnamefont {Sakai}},\ }\bibfield
  {title} {\bibinfo {title} {{Particle injection into the wave acceleration
  phase due to nonlinear wake wave breaking}},\ }\href@noop {} {\bibfield
  {journal} {\bibinfo  {journal} {Physical Review E}\ }\textbf {\bibinfo
  {volume} {58}},\ \bibinfo {pages} {R5257} (\bibinfo {year}
  {1998})}\BibitemShut {NoStop}%
\bibitem [{\citenamefont {Suk}\ \emph {et~al.}(2001)\citenamefont {Suk},
  \citenamefont {Barov}, \citenamefont {Rosenzweig},\ and\ \citenamefont
  {Esarey}}]{Suk2001}%
  \BibitemOpen
  \bibfield  {author} {\bibinfo {author} {\bibfnamefont {H.}~\bibnamefont
  {Suk}}, \bibinfo {author} {\bibfnamefont {N.}~\bibnamefont {Barov}}, \bibinfo
  {author} {\bibfnamefont {J.~B.}\ \bibnamefont {Rosenzweig}},\ and\ \bibinfo
  {author} {\bibfnamefont {E.}~\bibnamefont {Esarey}},\ }\bibfield  {title}
  {\bibinfo {title} {{Plasma electron trapping and acceleration in a plasma
  wake field using a density transition}},\ }\href@noop {} {\bibfield
  {journal} {\bibinfo  {journal} {Physical Review Letters}\ }\textbf {\bibinfo
  {volume} {86}},\ \bibinfo {pages} {1011} (\bibinfo {year}
  {2001})}\BibitemShut {NoStop}%
\bibitem [{\citenamefont {Buck}\ \emph {et~al.}(2013)\citenamefont {Buck},
  \citenamefont {Wenz}, \citenamefont {Xu}, \citenamefont {Khrennikov},
  \citenamefont {Schmid}, \citenamefont {Heigoldt}, \citenamefont {Mikhailova},
  \citenamefont {Geissler}, \citenamefont {Shen}, \citenamefont {Krausz},
  \citenamefont {Karsch},\ and\ \citenamefont {Veisz}}]{Buck2013}%
  \BibitemOpen
  \bibfield  {author} {\bibinfo {author} {\bibfnamefont {A.}~\bibnamefont
  {Buck}}, \bibinfo {author} {\bibfnamefont {J.}~\bibnamefont {Wenz}}, \bibinfo
  {author} {\bibfnamefont {J.}~\bibnamefont {Xu}}, \bibinfo {author}
  {\bibfnamefont {K.}~\bibnamefont {Khrennikov}}, \bibinfo {author}
  {\bibfnamefont {K.}~\bibnamefont {Schmid}}, \bibinfo {author} {\bibfnamefont
  {M.}~\bibnamefont {Heigoldt}}, \bibinfo {author} {\bibfnamefont {J.~M.}\
  \bibnamefont {Mikhailova}}, \bibinfo {author} {\bibfnamefont
  {M.}~\bibnamefont {Geissler}}, \bibinfo {author} {\bibfnamefont
  {B.}~\bibnamefont {Shen}}, \bibinfo {author} {\bibfnamefont {F.}~\bibnamefont
  {Krausz}}, \bibinfo {author} {\bibfnamefont {S.}~\bibnamefont {Karsch}},\
  and\ \bibinfo {author} {\bibfnamefont {L.}~\bibnamefont {Veisz}},\ }\bibfield
   {title} {\bibinfo {title} {{Shock-Front Injector for High-Quality
  Laser-Plasma Acceleration}},\ }\href@noop {} {\bibfield  {journal} {\bibinfo
  {journal} {Physical Review Letters}\ }\textbf {\bibinfo {volume} {110}},\
  \bibinfo {pages} {185006} (\bibinfo {year} {2013})}\BibitemShut {NoStop}%
\bibitem [{\citenamefont {Geddes}\ \emph {et~al.}(2008)\citenamefont {Geddes},
  \citenamefont {Nakamura}, \citenamefont {Plateau}, \citenamefont {Toth},
  \citenamefont {Cormier-Michel}, \citenamefont {Esarey}, \citenamefont
  {Schroeder}, \citenamefont {Cary},\ and\ \citenamefont
  {Leemans}}]{Geddes2008}%
  \BibitemOpen
  \bibfield  {author} {\bibinfo {author} {\bibfnamefont {C.~G.~R.}\
  \bibnamefont {Geddes}}, \bibinfo {author} {\bibfnamefont {K.}~\bibnamefont
  {Nakamura}}, \bibinfo {author} {\bibfnamefont {G.~R.}\ \bibnamefont
  {Plateau}}, \bibinfo {author} {\bibfnamefont {C.}~\bibnamefont {Toth}},
  \bibinfo {author} {\bibfnamefont {E.}~\bibnamefont {Cormier-Michel}},
  \bibinfo {author} {\bibfnamefont {E.}~\bibnamefont {Esarey}}, \bibinfo
  {author} {\bibfnamefont {C.~B.}\ \bibnamefont {Schroeder}}, \bibinfo {author}
  {\bibfnamefont {J.~R.}\ \bibnamefont {Cary}},\ and\ \bibinfo {author}
  {\bibfnamefont {W.~P.}\ \bibnamefont {Leemans}},\ }\bibfield  {title}
  {\bibinfo {title} {{Plasma-density-gradient injection of low
  absolute-momentum-spread electron bunches}},\ }\href@noop {} {\bibfield
  {journal} {\bibinfo  {journal} {Physical Review Letters}\ }\textbf {\bibinfo
  {volume} {100}},\ \bibinfo {pages} {215004} (\bibinfo {year}
  {2008})}\BibitemShut {NoStop}%
\bibitem [{\citenamefont {Xu}\ \emph {et~al.}(2017)\citenamefont {Xu},
  \citenamefont {Li}, \citenamefont {An}, \citenamefont {Dalichaouch},
  \citenamefont {Yu}, \citenamefont {Lu}, \citenamefont {Joshi},\ and\
  \citenamefont {Mori}}]{Xu2017}%
  \BibitemOpen
  \bibfield  {author} {\bibinfo {author} {\bibfnamefont {X.~L.}\ \bibnamefont
  {Xu}}, \bibinfo {author} {\bibfnamefont {F.}~\bibnamefont {Li}}, \bibinfo
  {author} {\bibfnamefont {W.}~\bibnamefont {An}}, \bibinfo {author}
  {\bibfnamefont {T.~N.}\ \bibnamefont {Dalichaouch}}, \bibinfo {author}
  {\bibfnamefont {P.}~\bibnamefont {Yu}}, \bibinfo {author} {\bibfnamefont
  {W.}~\bibnamefont {Lu}}, \bibinfo {author} {\bibfnamefont {C.}~\bibnamefont
  {Joshi}},\ and\ \bibinfo {author} {\bibfnamefont {W.~B.}\ \bibnamefont
  {Mori}},\ }\bibfield  {title} {\bibinfo {title} {{High quality electron bunch
  generation using a longitudinal density-tailored plasma-based accelerator in
  the three-dimensional blowout regime}},\ }\href@noop {} {\bibfield  {journal}
  {\bibinfo  {journal} {Physical Review Accelerators and Beams}\ }\textbf
  {\bibinfo {volume} {20}},\ \bibinfo {pages} {111303} (\bibinfo {year}
  {2017})}\BibitemShut {NoStop}%
\bibitem [{\citenamefont {Kalmykov}\ \emph {et~al.}(2009)\citenamefont
  {Kalmykov}, \citenamefont {Yi}, \citenamefont {Khudik},\ and\ \citenamefont
  {Shvets}}]{Kalmykov2009}%
  \BibitemOpen
  \bibfield  {author} {\bibinfo {author} {\bibfnamefont {S.}~\bibnamefont
  {Kalmykov}}, \bibinfo {author} {\bibfnamefont {S.~A.}\ \bibnamefont {Yi}},
  \bibinfo {author} {\bibfnamefont {V.}~\bibnamefont {Khudik}},\ and\ \bibinfo
  {author} {\bibfnamefont {G.}~\bibnamefont {Shvets}},\ }\bibfield  {title}
  {\bibinfo {title} {{Electron self-injection and trapping into an evolving
  plasma bubble}},\ }\href@noop {} {\bibfield  {journal} {\bibinfo  {journal}
  {Physical Review Letters}\ }\textbf {\bibinfo {volume} {103}},\ \bibinfo
  {pages} {135004} (\bibinfo {year} {2009})}\BibitemShut {NoStop}%
\bibitem [{\citenamefont {Lehe}\ \emph {et~al.}(2013)\citenamefont {Lehe},
  \citenamefont {Lifschitz}, \citenamefont {Davoine}, \citenamefont {Thaury},\
  and\ \citenamefont {Malka}}]{Lehe2013}%
  \BibitemOpen
  \bibfield  {author} {\bibinfo {author} {\bibfnamefont {R.}~\bibnamefont
  {Lehe}}, \bibinfo {author} {\bibfnamefont {A.~F.}\ \bibnamefont {Lifschitz}},
  \bibinfo {author} {\bibfnamefont {X.}~\bibnamefont {Davoine}}, \bibinfo
  {author} {\bibfnamefont {C.}~\bibnamefont {Thaury}},\ and\ \bibinfo {author}
  {\bibfnamefont {V.}~\bibnamefont {Malka}},\ }\bibfield  {title} {\bibinfo
  {title} {{Optical transverse injection in laser-plasma acceleration}},\
  }\href@noop {} {\bibfield  {journal} {\bibinfo  {journal} {Physical Review
  Letters}\ }\textbf {\bibinfo {volume} {111}},\ \bibinfo {pages} {085005}
  (\bibinfo {year} {2013})}\BibitemShut {NoStop}%
\bibitem [{\citenamefont {Dalichaouch}\ \emph {et~al.}(2020)\citenamefont
  {Dalichaouch}, \citenamefont {Xu}, \citenamefont {Li}, \citenamefont
  {Tableman}, \citenamefont {Tsung}, \citenamefont {An},\ and\ \citenamefont
  {Mori}}]{Dalichaouch2020}%
  \BibitemOpen
  \bibfield  {author} {\bibinfo {author} {\bibfnamefont {T.~N.}\ \bibnamefont
  {Dalichaouch}}, \bibinfo {author} {\bibfnamefont {X.~L.}\ \bibnamefont {Xu}},
  \bibinfo {author} {\bibfnamefont {F.}~\bibnamefont {Li}}, \bibinfo {author}
  {\bibfnamefont {A.}~\bibnamefont {Tableman}}, \bibinfo {author}
  {\bibfnamefont {F.~S.}\ \bibnamefont {Tsung}}, \bibinfo {author}
  {\bibfnamefont {W.}~\bibnamefont {An}},\ and\ \bibinfo {author}
  {\bibfnamefont {W.}~\bibnamefont {Mori}},\ }\bibfield  {title} {\bibinfo
  {title} {{Generating high quality ultrarelativistic electron beams using an
  evolving electron beam driver}},\ }\href@noop {} {\bibfield  {journal}
  {\bibinfo  {journal} {Physical Review Accelerators and Beams}\ }\textbf
  {\bibinfo {volume} {23}},\ \bibinfo {pages} {021304} (\bibinfo {year}
  {2020})}\BibitemShut {NoStop}%
\bibitem [{\citenamefont {Rosenzweig}\ \emph {et~al.}(1991)\citenamefont
  {Rosenzweig}, \citenamefont {Breizman}, \citenamefont {Katsouleas},\ and\
  \citenamefont {Su}}]{Rosenzweig1991}%
  \BibitemOpen
  \bibfield  {author} {\bibinfo {author} {\bibfnamefont {J.~B.}\ \bibnamefont
  {Rosenzweig}}, \bibinfo {author} {\bibfnamefont {B.}~\bibnamefont
  {Breizman}}, \bibinfo {author} {\bibfnamefont {T.}~\bibnamefont
  {Katsouleas}},\ and\ \bibinfo {author} {\bibfnamefont {J.~J.}\ \bibnamefont
  {Su}},\ }\bibfield  {title} {\bibinfo {title} {{Acceleration and focusing of
  electrons in two-dimensional nonlinear plasma wake fields}},\ }\href@noop {}
  {\bibfield  {journal} {\bibinfo  {journal} {Physical Review A}\ }\textbf
  {\bibinfo {volume} {44}},\ \bibinfo {pages} {R6189} (\bibinfo {year}
  {1991})}\BibitemShut {NoStop}%
\bibitem [{\citenamefont {Pukhov}\ and\ \citenamefont {Meyer-ter
  Vehn}(2002)}]{Pukhov2002}%
  \BibitemOpen
  \bibfield  {author} {\bibinfo {author} {\bibfnamefont {A.}~\bibnamefont
  {Pukhov}}\ and\ \bibinfo {author} {\bibfnamefont {J.}~\bibnamefont {Meyer-ter
  Vehn}},\ }\bibfield  {title} {\bibinfo {title} {{Laser wake field
  acceleration: the highly non-linear broken-wave regime}},\ }\href@noop {}
  {\bibfield  {journal} {\bibinfo  {journal} {Applied Physics B}\ }\textbf
  {\bibinfo {volume} {74}},\ \bibinfo {pages} {355} (\bibinfo {year}
  {2002})}\BibitemShut {NoStop}%
\bibitem [{\citenamefont {Lu}\ \emph {et~al.}(2006{\natexlab{a}})\citenamefont
  {Lu}, \citenamefont {Huang}, \citenamefont {Zhou}, \citenamefont {Mori},\
  and\ \citenamefont {Katsouleas}}]{Lu2006a}%
  \BibitemOpen
  \bibfield  {author} {\bibinfo {author} {\bibfnamefont {W.}~\bibnamefont
  {Lu}}, \bibinfo {author} {\bibfnamefont {C.}~\bibnamefont {Huang}}, \bibinfo
  {author} {\bibfnamefont {M.}~\bibnamefont {Zhou}}, \bibinfo {author}
  {\bibfnamefont {W.~B.}\ \bibnamefont {Mori}},\ and\ \bibinfo {author}
  {\bibfnamefont {T.}~\bibnamefont {Katsouleas}},\ }\bibfield  {title}
  {\bibinfo {title} {{Nonlinear theory for relativistic plasma wakefields in
  the blowout regime}},\ }\href@noop {} {\bibfield  {journal} {\bibinfo
  {journal} {Physical Review Letters}\ }\textbf {\bibinfo {volume} {96}},\
  \bibinfo {pages} {165002} (\bibinfo {year} {2006}{\natexlab{a}})}\BibitemShut
  {NoStop}%
\bibitem [{\citenamefont {Lu}\ \emph {et~al.}(2006{\natexlab{b}})\citenamefont
  {Lu}, \citenamefont {Huang}, \citenamefont {Zhou}, \citenamefont {Tzoufras},
  \citenamefont {Tsung}, \citenamefont {Mori},\ and\ \citenamefont
  {Katsouleas}}]{Lu2006b}%
  \BibitemOpen
  \bibfield  {author} {\bibinfo {author} {\bibfnamefont {W.}~\bibnamefont
  {Lu}}, \bibinfo {author} {\bibfnamefont {C.}~\bibnamefont {Huang}}, \bibinfo
  {author} {\bibfnamefont {M.}~\bibnamefont {Zhou}}, \bibinfo {author}
  {\bibfnamefont {M.}~\bibnamefont {Tzoufras}}, \bibinfo {author}
  {\bibfnamefont {F.~S.}\ \bibnamefont {Tsung}}, \bibinfo {author}
  {\bibfnamefont {W.~B.}\ \bibnamefont {Mori}},\ and\ \bibinfo {author}
  {\bibfnamefont {T.}~\bibnamefont {Katsouleas}},\ }\bibfield  {title}
  {\bibinfo {title} {{A nonlinear theory for multidimensional relativistic
  plasma wave wakefields}},\ }\href@noop {} {\bibfield  {journal} {\bibinfo
  {journal} {Physics of Plasmas}\ }\textbf {\bibinfo {volume} {13}},\ \bibinfo
  {pages} {056709} (\bibinfo {year} {2006}{\natexlab{b}})}\BibitemShut
  {NoStop}%
\bibitem [{\citenamefont {Froula}\ \emph {et~al.}(2018)\citenamefont {Froula},
  \citenamefont {Turnbull}, \citenamefont {Davies}, \citenamefont {Kessler},
  \citenamefont {Haberberger}, \citenamefont {Palastro}, \citenamefont {Bahk},
  \citenamefont {Begishev}, \citenamefont {Boni}, \citenamefont {Bucht},
  \citenamefont {Katz},\ and\ \citenamefont {Shaw}}]{Froula2018}%
  \BibitemOpen
  \bibfield  {author} {\bibinfo {author} {\bibfnamefont {D.~H.}\ \bibnamefont
  {Froula}}, \bibinfo {author} {\bibfnamefont {D.}~\bibnamefont {Turnbull}},
  \bibinfo {author} {\bibfnamefont {A.~S.}\ \bibnamefont {Davies}}, \bibinfo
  {author} {\bibfnamefont {T.~J.}\ \bibnamefont {Kessler}}, \bibinfo {author}
  {\bibfnamefont {D.}~\bibnamefont {Haberberger}}, \bibinfo {author}
  {\bibfnamefont {J.~P.}\ \bibnamefont {Palastro}}, \bibinfo {author}
  {\bibfnamefont {S.~W.}\ \bibnamefont {Bahk}}, \bibinfo {author}
  {\bibfnamefont {I.~A.}\ \bibnamefont {Begishev}}, \bibinfo {author}
  {\bibfnamefont {R.}~\bibnamefont {Boni}}, \bibinfo {author} {\bibfnamefont
  {S.}~\bibnamefont {Bucht}}, \bibinfo {author} {\bibfnamefont
  {J.}~\bibnamefont {Katz}},\ and\ \bibinfo {author} {\bibfnamefont {J.~L.}\
  \bibnamefont {Shaw}},\ }\bibfield  {title} {\bibinfo {title} {{Spatiotemporal
  control of laser intensity}},\ }\href@noop {} {\bibfield  {journal} {\bibinfo
   {journal} {Nature Photonics}\ }\textbf {\bibinfo {volume} {12}},\ \bibinfo
  {pages} {262} (\bibinfo {year} {2018})}\BibitemShut {NoStop}%
\bibitem [{\citenamefont {Palastro}\ \emph {et~al.}(2018)\citenamefont
  {Palastro}, \citenamefont {Turnbull}, \citenamefont {Bahk}, \citenamefont
  {Follett}, \citenamefont {Shaw}, \citenamefont {Haberberger}, \citenamefont
  {Bromage},\ and\ \citenamefont {Froula}}]{Palastro2018}%
  \BibitemOpen
  \bibfield  {author} {\bibinfo {author} {\bibfnamefont {J.~P.}\ \bibnamefont
  {Palastro}}, \bibinfo {author} {\bibfnamefont {D.}~\bibnamefont {Turnbull}},
  \bibinfo {author} {\bibfnamefont {S.~W.}\ \bibnamefont {Bahk}}, \bibinfo
  {author} {\bibfnamefont {R.~K.}\ \bibnamefont {Follett}}, \bibinfo {author}
  {\bibfnamefont {J.~L.}\ \bibnamefont {Shaw}}, \bibinfo {author}
  {\bibfnamefont {D.}~\bibnamefont {Haberberger}}, \bibinfo {author}
  {\bibfnamefont {J.}~\bibnamefont {Bromage}},\ and\ \bibinfo {author}
  {\bibfnamefont {D.~H.}\ \bibnamefont {Froula}},\ }\bibfield  {title}
  {\bibinfo {title} {{Ionization waves of arbitrary velocity driven by a flying
  focus}},\ }\href@noop {} {\bibfield  {journal} {\bibinfo  {journal} {Physical
  Review A}\ }\textbf {\bibinfo {volume} {97}},\ \bibinfo {pages} {033835}
  (\bibinfo {year} {2018})}\BibitemShut {NoStop}%
\bibitem [{\citenamefont {Kondakci}\ and\ \citenamefont
  {Abouraddy}(2019)}]{Kondakci2019}%
  \BibitemOpen
  \bibfield  {author} {\bibinfo {author} {\bibfnamefont {H.~E.}\ \bibnamefont
  {Kondakci}}\ and\ \bibinfo {author} {\bibfnamefont {A.~F.}\ \bibnamefont
  {Abouraddy}},\ }\bibfield  {title} {\bibinfo {title} {{Optical space-time
  wave packets having arbitrary group velocities in free space}},\ }\href@noop
  {} {\bibfield  {journal} {\bibinfo  {journal} {Nature Communications}\
  }\textbf {\bibinfo {volume} {10}},\ \bibinfo {pages} {1} (\bibinfo {year}
  {2019})}\BibitemShut {NoStop}%
\bibitem [{\citenamefont {Palastro}\ \emph {et~al.}(2020)\citenamefont
  {Palastro}, \citenamefont {Shaw}, \citenamefont {Franke}, \citenamefont
  {Ramsey}, \citenamefont {Simpson},\ and\ \citenamefont
  {Froula}}]{Palastro2020}%
  \BibitemOpen
  \bibfield  {author} {\bibinfo {author} {\bibfnamefont {J.~P.}\ \bibnamefont
  {Palastro}}, \bibinfo {author} {\bibfnamefont {J.~L.}\ \bibnamefont {Shaw}},
  \bibinfo {author} {\bibfnamefont {P.}~\bibnamefont {Franke}}, \bibinfo
  {author} {\bibfnamefont {D.}~\bibnamefont {Ramsey}}, \bibinfo {author}
  {\bibfnamefont {T.~T.}\ \bibnamefont {Simpson}},\ and\ \bibinfo {author}
  {\bibfnamefont {D.~H.}\ \bibnamefont {Froula}},\ }\bibfield  {title}
  {\bibinfo {title} {{Dephasingless Laser Wakefield Acceleration}},\
  }\href@noop {} {\bibfield  {journal} {\bibinfo  {journal} {Physical Review
  Letters}\ }\textbf {\bibinfo {volume} {124}},\ \bibinfo {pages} {134802}
  (\bibinfo {year} {2020})}\BibitemShut {NoStop}%
\bibitem [{\citenamefont {Yessenov}\ and\ \citenamefont
  {Abouraddy}(2020)}]{Yessenov2020}%
  \BibitemOpen
  \bibfield  {author} {\bibinfo {author} {\bibfnamefont {M.}~\bibnamefont
  {Yessenov}}\ and\ \bibinfo {author} {\bibfnamefont {A.~F.}\ \bibnamefont
  {Abouraddy}},\ }\bibfield  {title} {\bibinfo {title} {{Accelerating and
  Decelerating Space-Time Optical Wave Packets in Free Space}},\ }\href@noop {}
  {\bibfield  {journal} {\bibinfo  {journal} {Physical Review Letters}\
  }\textbf {\bibinfo {volume} {125}},\ \bibinfo {pages} {233901} (\bibinfo
  {year} {2020})}\BibitemShut {NoStop}%
\bibitem [{\citenamefont {Caizergues}\ \emph {et~al.}(2020)\citenamefont
  {Caizergues}, \citenamefont {Smartsev}, \citenamefont {Malka},\ and\
  \citenamefont {Thaury}}]{Caizergues2020}%
  \BibitemOpen
  \bibfield  {author} {\bibinfo {author} {\bibfnamefont {C.}~\bibnamefont
  {Caizergues}}, \bibinfo {author} {\bibfnamefont {S.}~\bibnamefont
  {Smartsev}}, \bibinfo {author} {\bibfnamefont {V.}~\bibnamefont {Malka}},\
  and\ \bibinfo {author} {\bibfnamefont {C.}~\bibnamefont {Thaury}},\
  }\bibfield  {title} {\bibinfo {title} {{Phase-locked laser-wakefield electron
  acceleration}},\ }\href@noop {} {\bibfield  {journal} {\bibinfo  {journal}
  {Nature Photonics}\ }\textbf {\bibinfo {volume} {14}},\ \bibinfo {pages}
  {475} (\bibinfo {year} {2020})}\BibitemShut {NoStop}%
\bibitem [{\citenamefont {Hogan}\ \emph {et~al.}(2005)\citenamefont {Hogan},
  \citenamefont {Barnes}, \citenamefont {Clayton}, \citenamefont {Decker},
  \citenamefont {Deng}, \citenamefont {Emma}, \citenamefont {Huang},
  \citenamefont {Iverson}, \citenamefont {Johnson}, \citenamefont {Joshi},
  \citenamefont {Katsouleas}, \citenamefont {Krejcik}, \citenamefont {Lu},
  \citenamefont {Marsh}, \citenamefont {Mori}, \citenamefont {Muggli},
  \citenamefont {O'Connell}, \citenamefont {Oz}, \citenamefont {Siemann},\ and\
  \citenamefont {Walz}}]{Hogan2005}%
  \BibitemOpen
  \bibfield  {author} {\bibinfo {author} {\bibfnamefont {M.~J.}\ \bibnamefont
  {Hogan}}, \bibinfo {author} {\bibfnamefont {C.~D.}\ \bibnamefont {Barnes}},
  \bibinfo {author} {\bibfnamefont {C.~E.}\ \bibnamefont {Clayton}}, \bibinfo
  {author} {\bibfnamefont {F.~J.}\ \bibnamefont {Decker}}, \bibinfo {author}
  {\bibfnamefont {S.}~\bibnamefont {Deng}}, \bibinfo {author} {\bibfnamefont
  {P.}~\bibnamefont {Emma}}, \bibinfo {author} {\bibfnamefont {C.}~\bibnamefont
  {Huang}}, \bibinfo {author} {\bibfnamefont {R.~H.}\ \bibnamefont {Iverson}},
  \bibinfo {author} {\bibfnamefont {D.~K.}\ \bibnamefont {Johnson}}, \bibinfo
  {author} {\bibfnamefont {C.}~\bibnamefont {Joshi}}, \bibinfo {author}
  {\bibfnamefont {T.}~\bibnamefont {Katsouleas}}, \bibinfo {author}
  {\bibfnamefont {P.}~\bibnamefont {Krejcik}}, \bibinfo {author} {\bibfnamefont
  {W.}~\bibnamefont {Lu}}, \bibinfo {author} {\bibfnamefont {K.~A.}\
  \bibnamefont {Marsh}}, \bibinfo {author} {\bibfnamefont {W.~B.}\ \bibnamefont
  {Mori}}, \bibinfo {author} {\bibfnamefont {P.}~\bibnamefont {Muggli}},
  \bibinfo {author} {\bibfnamefont {C.~L.}\ \bibnamefont {O'Connell}}, \bibinfo
  {author} {\bibfnamefont {E.}~\bibnamefont {Oz}}, \bibinfo {author}
  {\bibfnamefont {R.~H.}\ \bibnamefont {Siemann}},\ and\ \bibinfo {author}
  {\bibfnamefont {D.}~\bibnamefont {Walz}},\ }\bibfield  {title} {\bibinfo
  {title} {{Multi-GeV Energy Gain in a Plasma-Wakefield Accelerator}},\
  }\href@noop {} {\bibfield  {journal} {\bibinfo  {journal} {Physical Review
  Letters}\ }\textbf {\bibinfo {volume} {95}},\ \bibinfo {pages} {054802}
  (\bibinfo {year} {2005})}\BibitemShut {NoStop}%
\bibitem [{\citenamefont {Litos}\ \emph {et~al.}(2014)\citenamefont {Litos},
  \citenamefont {Adli}, \citenamefont {An}, \citenamefont {Clarke},
  \citenamefont {Clayton}, \citenamefont {Corde}, \citenamefont {Delahaye},
  \citenamefont {England}, \citenamefont {Fisher}, \citenamefont {Frederico},
  \citenamefont {Gessner}, \citenamefont {Green}, \citenamefont {Hogan},
  \citenamefont {Joshi}, \citenamefont {Lu}, \citenamefont {Marsh},
  \citenamefont {Mori}, \citenamefont {Muggli}, \citenamefont
  {Vafaei-Najafabadi}, \citenamefont {Walz}, \citenamefont {White},
  \citenamefont {Wu}, \citenamefont {Yakimenko},\ and\ \citenamefont
  {Yocky}}]{Litos2014}%
  \BibitemOpen
  \bibfield  {author} {\bibinfo {author} {\bibfnamefont {M.}~\bibnamefont
  {Litos}}, \bibinfo {author} {\bibfnamefont {E.}~\bibnamefont {Adli}},
  \bibinfo {author} {\bibfnamefont {W.}~\bibnamefont {An}}, \bibinfo {author}
  {\bibfnamefont {C.~I.}\ \bibnamefont {Clarke}}, \bibinfo {author}
  {\bibfnamefont {C.~E.}\ \bibnamefont {Clayton}}, \bibinfo {author}
  {\bibfnamefont {S.}~\bibnamefont {Corde}}, \bibinfo {author} {\bibfnamefont
  {J.~P.}\ \bibnamefont {Delahaye}}, \bibinfo {author} {\bibfnamefont {R.~J.}\
  \bibnamefont {England}}, \bibinfo {author} {\bibfnamefont {A.~S.}\
  \bibnamefont {Fisher}}, \bibinfo {author} {\bibfnamefont {J.}~\bibnamefont
  {Frederico}}, \bibinfo {author} {\bibfnamefont {S.}~\bibnamefont {Gessner}},
  \bibinfo {author} {\bibfnamefont {S.~Z.}\ \bibnamefont {Green}}, \bibinfo
  {author} {\bibfnamefont {M.~J.}\ \bibnamefont {Hogan}}, \bibinfo {author}
  {\bibfnamefont {C.}~\bibnamefont {Joshi}}, \bibinfo {author} {\bibfnamefont
  {W.}~\bibnamefont {Lu}}, \bibinfo {author} {\bibfnamefont {K.~A.}\
  \bibnamefont {Marsh}}, \bibinfo {author} {\bibfnamefont {W.~B.}\ \bibnamefont
  {Mori}}, \bibinfo {author} {\bibfnamefont {P.}~\bibnamefont {Muggli}},
  \bibinfo {author} {\bibfnamefont {N.}~\bibnamefont {Vafaei-Najafabadi}},
  \bibinfo {author} {\bibfnamefont {D.}~\bibnamefont {Walz}}, \bibinfo {author}
  {\bibfnamefont {G.}~\bibnamefont {White}}, \bibinfo {author} {\bibfnamefont
  {Z.}~\bibnamefont {Wu}}, \bibinfo {author} {\bibfnamefont {V.}~\bibnamefont
  {Yakimenko}},\ and\ \bibinfo {author} {\bibfnamefont {G.}~\bibnamefont
  {Yocky}},\ }\bibfield  {title} {\bibinfo {title} {{High-efficiency
  acceleration of an electron beam in a plasma wakefield accelerator}},\
  }\href@noop {} {\bibfield  {journal} {\bibinfo  {journal} {Nature}\ }\textbf
  {\bibinfo {volume} {515}},\ \bibinfo {pages} {92} (\bibinfo {year}
  {2014})}\BibitemShut {NoStop}%
\bibitem [{\citenamefont {Su}\ \emph {et~al.}(1990)\citenamefont {Su},
  \citenamefont {Katsouleas}, \citenamefont {Dawson},\ and\ \citenamefont
  {Fedele}}]{Su1990}%
  \BibitemOpen
  \bibfield  {author} {\bibinfo {author} {\bibfnamefont {J.~J.}\ \bibnamefont
  {Su}}, \bibinfo {author} {\bibfnamefont {T.}~\bibnamefont {Katsouleas}},
  \bibinfo {author} {\bibfnamefont {J.~M.}\ \bibnamefont {Dawson}},\ and\
  \bibinfo {author} {\bibfnamefont {R.}~\bibnamefont {Fedele}},\ }\bibfield
  {title} {\bibinfo {title} {{Plasma lenses for focusing particle beams}},\
  }\href@noop {} {\bibfield  {journal} {\bibinfo  {journal} {Phys. Rev. A}\
  }\textbf {\bibinfo {volume} {41}},\ \bibinfo {pages} {3321} (\bibinfo {year}
  {1990})}\BibitemShut {NoStop}%
\bibitem [{\citenamefont {Pompili}\ \emph {et~al.}(2018)\citenamefont
  {Pompili}, \citenamefont {Anania}, \citenamefont {Bellaveglia}, \citenamefont
  {Biagioni}, \citenamefont {Bini}, \citenamefont {Bisesto}, \citenamefont
  {Brentegani}, \citenamefont {Cardelli}, \citenamefont {Castorina},
  \citenamefont {Chiadroni}, \citenamefont {Cianchi}, \citenamefont {Coiro},
  \citenamefont {Costa}, \citenamefont {Croia}, \citenamefont {Di~Giovenale},
  \citenamefont {Ferrario}, \citenamefont {Filippi}, \citenamefont {Giribono},
  \citenamefont {Lollo}, \citenamefont {Marocchino}, \citenamefont {Marongiu},
  \citenamefont {Martinelli}, \citenamefont {Mostacci}, \citenamefont
  {Pellegrini}, \citenamefont {Piersanti}, \citenamefont {Di~Pirro},
  \citenamefont {Romeo}, \citenamefont {Rossi}, \citenamefont {Scifo},
  \citenamefont {Shpakov}, \citenamefont {Stella}, \citenamefont {Vaccarezza},
  \citenamefont {Villa},\ and\ \citenamefont {Zigler}}]{Pompili2018}%
  \BibitemOpen
  \bibfield  {author} {\bibinfo {author} {\bibfnamefont {R.}~\bibnamefont
  {Pompili}}, \bibinfo {author} {\bibfnamefont {M.~P.}\ \bibnamefont {Anania}},
  \bibinfo {author} {\bibfnamefont {M.}~\bibnamefont {Bellaveglia}}, \bibinfo
  {author} {\bibfnamefont {A.}~\bibnamefont {Biagioni}}, \bibinfo {author}
  {\bibfnamefont {S.}~\bibnamefont {Bini}}, \bibinfo {author} {\bibfnamefont
  {F.}~\bibnamefont {Bisesto}}, \bibinfo {author} {\bibfnamefont
  {E.}~\bibnamefont {Brentegani}}, \bibinfo {author} {\bibfnamefont
  {F.}~\bibnamefont {Cardelli}}, \bibinfo {author} {\bibfnamefont
  {G.}~\bibnamefont {Castorina}}, \bibinfo {author} {\bibfnamefont
  {E.}~\bibnamefont {Chiadroni}}, \bibinfo {author} {\bibfnamefont
  {A.}~\bibnamefont {Cianchi}}, \bibinfo {author} {\bibfnamefont
  {O.}~\bibnamefont {Coiro}}, \bibinfo {author} {\bibfnamefont
  {G.}~\bibnamefont {Costa}}, \bibinfo {author} {\bibfnamefont
  {M.}~\bibnamefont {Croia}}, \bibinfo {author} {\bibfnamefont
  {D.}~\bibnamefont {Di~Giovenale}}, \bibinfo {author} {\bibfnamefont
  {M.}~\bibnamefont {Ferrario}}, \bibinfo {author} {\bibfnamefont
  {F.}~\bibnamefont {Filippi}}, \bibinfo {author} {\bibfnamefont
  {A.}~\bibnamefont {Giribono}}, \bibinfo {author} {\bibfnamefont
  {V.}~\bibnamefont {Lollo}}, \bibinfo {author} {\bibfnamefont
  {A.}~\bibnamefont {Marocchino}}, \bibinfo {author} {\bibfnamefont
  {M.}~\bibnamefont {Marongiu}}, \bibinfo {author} {\bibfnamefont
  {V.}~\bibnamefont {Martinelli}}, \bibinfo {author} {\bibfnamefont
  {A.}~\bibnamefont {Mostacci}}, \bibinfo {author} {\bibfnamefont
  {D.}~\bibnamefont {Pellegrini}}, \bibinfo {author} {\bibfnamefont
  {L.}~\bibnamefont {Piersanti}}, \bibinfo {author} {\bibfnamefont
  {G.}~\bibnamefont {Di~Pirro}}, \bibinfo {author} {\bibfnamefont
  {S.}~\bibnamefont {Romeo}}, \bibinfo {author} {\bibfnamefont {A.~R.}\
  \bibnamefont {Rossi}}, \bibinfo {author} {\bibfnamefont {J.}~\bibnamefont
  {Scifo}}, \bibinfo {author} {\bibfnamefont {V.}~\bibnamefont {Shpakov}},
  \bibinfo {author} {\bibfnamefont {A.}~\bibnamefont {Stella}}, \bibinfo
  {author} {\bibfnamefont {C.}~\bibnamefont {Vaccarezza}}, \bibinfo {author}
  {\bibfnamefont {F.}~\bibnamefont {Villa}},\ and\ \bibinfo {author}
  {\bibfnamefont {A.}~\bibnamefont {Zigler}},\ }\bibfield  {title} {\bibinfo
  {title} {{Focusing of High-Brightness Electron Beams with Active-Plasma
  Lenses}},\ }\href@noop {} {\bibfield  {journal} {\bibinfo  {journal} {Phys.
  Rev. Lett.}\ }\textbf {\bibinfo {volume} {121}},\ \bibinfo {pages} {174801}
  (\bibinfo {year} {2018})}\BibitemShut {NoStop}%
\bibitem [{\citenamefont {Fonseca}\ \emph {et~al.}(2002)\citenamefont
  {Fonseca}, \citenamefont {Silva}, \citenamefont {Tsung}, \citenamefont
  {Decyk}, \citenamefont {Lu}, \citenamefont {Ren}, \citenamefont {Mori},
  \citenamefont {Deng}, \citenamefont {Lee}, \citenamefont {Katsouleas},\ and\
  \citenamefont {Adam}}]{Fonseca2002}%
  \BibitemOpen
  \bibfield  {author} {\bibinfo {author} {\bibfnamefont {R.~A.}\ \bibnamefont
  {Fonseca}}, \bibinfo {author} {\bibfnamefont {L.~O.}\ \bibnamefont {Silva}},
  \bibinfo {author} {\bibfnamefont {F.~S.}\ \bibnamefont {Tsung}}, \bibinfo
  {author} {\bibfnamefont {V.~K.}\ \bibnamefont {Decyk}}, \bibinfo {author}
  {\bibfnamefont {W.}~\bibnamefont {Lu}}, \bibinfo {author} {\bibfnamefont
  {C.}~\bibnamefont {Ren}}, \bibinfo {author} {\bibfnamefont {W.~B.}\
  \bibnamefont {Mori}}, \bibinfo {author} {\bibfnamefont {S.}~\bibnamefont
  {Deng}}, \bibinfo {author} {\bibfnamefont {S.}~\bibnamefont {Lee}}, \bibinfo
  {author} {\bibfnamefont {T.}~\bibnamefont {Katsouleas}},\ and\ \bibinfo
  {author} {\bibfnamefont {J.~C.}\ \bibnamefont {Adam}},\ }\bibfield  {title}
  {\bibinfo {title} {{OSIRIS: A three-dimensional, fully relativistic particle
  in cell code for modeling plasma based accelerators}},\ }in\ \href@noop {}
  {\emph {\bibinfo {booktitle} {Lecture Notes in Computer Science (including
  subseries Lecture Notes in Artificial Intelligence and Lecture Notes in
  Bioinformatics)}}},\ Vol.\ \bibinfo {volume} {2331 LNCS}\ (\bibinfo {year}
  {2002})\ pp.\ \bibinfo {pages} {342--351}\BibitemShut {NoStop}%
\bibitem [{\citenamefont {Li}\ \emph {et~al.}(2021)\citenamefont {Li},
  \citenamefont {Miller}, \citenamefont {Xu}, \citenamefont {Tsung},
  \citenamefont {Decyk}, \citenamefont {An}, \citenamefont {Fonseca},\ and\
  \citenamefont {Mori}}]{Li2021}%
  \BibitemOpen
  \bibfield  {author} {\bibinfo {author} {\bibfnamefont {F.}~\bibnamefont
  {Li}}, \bibinfo {author} {\bibfnamefont {K.~G.}\ \bibnamefont {Miller}},
  \bibinfo {author} {\bibfnamefont {X.}~\bibnamefont {Xu}}, \bibinfo {author}
  {\bibfnamefont {F.~S.}\ \bibnamefont {Tsung}}, \bibinfo {author}
  {\bibfnamefont {V.~K.}\ \bibnamefont {Decyk}}, \bibinfo {author}
  {\bibfnamefont {W.}~\bibnamefont {An}}, \bibinfo {author} {\bibfnamefont
  {R.~A.}\ \bibnamefont {Fonseca}},\ and\ \bibinfo {author} {\bibfnamefont
  {W.~B.}\ \bibnamefont {Mori}},\ }\bibfield  {title} {\bibinfo {title} {{A new
  field solver for modeling of relativistic particle-laser interactions using
  the particle-in-cell algorithm}},\ }\href@noop {} {\bibfield  {journal}
  {\bibinfo  {journal} {Computer Physics Communications}\ }\textbf {\bibinfo
  {volume} {258}},\ \bibinfo {pages} {107580} (\bibinfo {year}
  {2021})}\BibitemShut {NoStop}%
\bibitem [{\citenamefont {Li}\ \emph {et~al.}(2017)\citenamefont {Li},
  \citenamefont {Yu}, \citenamefont {Xu}, \citenamefont {Fiuza}, \citenamefont
  {Decyk}, \citenamefont {Dalichaouch}, \citenamefont {Davidson}, \citenamefont
  {Tableman}, \citenamefont {An}, \citenamefont {Tsung}, \citenamefont
  {Fonseca}, \citenamefont {Lu},\ and\ \citenamefont {Mori}}]{Li2017}%
  \BibitemOpen
  \bibfield  {author} {\bibinfo {author} {\bibfnamefont {F.}~\bibnamefont
  {Li}}, \bibinfo {author} {\bibfnamefont {P.}~\bibnamefont {Yu}}, \bibinfo
  {author} {\bibfnamefont {X.}~\bibnamefont {Xu}}, \bibinfo {author}
  {\bibfnamefont {F.}~\bibnamefont {Fiuza}}, \bibinfo {author} {\bibfnamefont
  {V.~K.}\ \bibnamefont {Decyk}}, \bibinfo {author} {\bibfnamefont
  {T.}~\bibnamefont {Dalichaouch}}, \bibinfo {author} {\bibfnamefont
  {A.}~\bibnamefont {Davidson}}, \bibinfo {author} {\bibfnamefont
  {A.}~\bibnamefont {Tableman}}, \bibinfo {author} {\bibfnamefont
  {W.}~\bibnamefont {An}}, \bibinfo {author} {\bibfnamefont {F.~S.}\
  \bibnamefont {Tsung}}, \bibinfo {author} {\bibfnamefont {R.~A.}\ \bibnamefont
  {Fonseca}}, \bibinfo {author} {\bibfnamefont {W.}~\bibnamefont {Lu}},\ and\
  \bibinfo {author} {\bibfnamefont {W.~B.}\ \bibnamefont {Mori}},\ }\bibfield
  {title} {\bibinfo {title} {{Controlling the numerical Cerenkov instability in
  PIC simulations using a customized finite difference Maxwell solver and a
  local FFT based current correction}},\ }\href@noop {} {\bibfield  {journal}
  {\bibinfo  {journal} {Computer Physics Communications}\ }\textbf {\bibinfo
  {volume} {214}},\ \bibinfo {pages} {6} (\bibinfo {year} {2017})}\BibitemShut
  {NoStop}%
\bibitem [{\citenamefont {Xu}\ \emph {et~al.}(2013)\citenamefont {Xu},
  \citenamefont {Yu}, \citenamefont {Martins}, \citenamefont {Tsung},
  \citenamefont {Decyk}, \citenamefont {Vieira}, \citenamefont {Fonseca},
  \citenamefont {Lu}, \citenamefont {Silva},\ and\ \citenamefont
  {Mori}}]{Xu2013}%
  \BibitemOpen
  \bibfield  {author} {\bibinfo {author} {\bibfnamefont {X.}~\bibnamefont
  {Xu}}, \bibinfo {author} {\bibfnamefont {P.}~\bibnamefont {Yu}}, \bibinfo
  {author} {\bibfnamefont {S.~F.}\ \bibnamefont {Martins}}, \bibinfo {author}
  {\bibfnamefont {F.~S.}\ \bibnamefont {Tsung}}, \bibinfo {author}
  {\bibfnamefont {V.~K.}\ \bibnamefont {Decyk}}, \bibinfo {author}
  {\bibfnamefont {J.}~\bibnamefont {Vieira}}, \bibinfo {author} {\bibfnamefont
  {R.~A.}\ \bibnamefont {Fonseca}}, \bibinfo {author} {\bibfnamefont
  {W.}~\bibnamefont {Lu}}, \bibinfo {author} {\bibfnamefont {L.~O.}\
  \bibnamefont {Silva}},\ and\ \bibinfo {author} {\bibfnamefont {W.~B.}\
  \bibnamefont {Mori}},\ }\bibfield  {title} {\bibinfo {title} {{Numerical
  instability due to relativistic plasma drift in EM-PIC simulations}},\
  }\href@noop {} {\bibfield  {journal} {\bibinfo  {journal} {Computer Physics
  Communications}\ }\textbf {\bibinfo {volume} {184}},\ \bibinfo {pages} {2503}
  (\bibinfo {year} {2013})}\BibitemShut {NoStop}%
\bibitem [{\citenamefont {Xu}\ \emph {et~al.}(2020{\natexlab{b}})\citenamefont
  {Xu}, \citenamefont {Li}, \citenamefont {Tsung}, \citenamefont {Dalichaouch},
  \citenamefont {An}, \citenamefont {Wen}, \citenamefont {Decyk}, \citenamefont
  {Fonseca}, \citenamefont {Hogan},\ and\ \citenamefont {Mori}}]{Xu2020}%
  \BibitemOpen
  \bibfield  {author} {\bibinfo {author} {\bibfnamefont {X.}~\bibnamefont
  {Xu}}, \bibinfo {author} {\bibfnamefont {F.}~\bibnamefont {Li}}, \bibinfo
  {author} {\bibfnamefont {F.~S.}\ \bibnamefont {Tsung}}, \bibinfo {author}
  {\bibfnamefont {T.~N.}\ \bibnamefont {Dalichaouch}}, \bibinfo {author}
  {\bibfnamefont {W.}~\bibnamefont {An}}, \bibinfo {author} {\bibfnamefont
  {H.}~\bibnamefont {Wen}}, \bibinfo {author} {\bibfnamefont {V.~K.}\
  \bibnamefont {Decyk}}, \bibinfo {author} {\bibfnamefont {R.~A.}\ \bibnamefont
  {Fonseca}}, \bibinfo {author} {\bibfnamefont {M.~J.}\ \bibnamefont {Hogan}},\
  and\ \bibinfo {author} {\bibfnamefont {W.~B.}\ \bibnamefont {Mori}},\
  }\bibfield  {title} {\bibinfo {title} {{On numerical errors to the fields
  surrounding a relativistically moving particle in PIC codes}},\ }\href@noop
  {} {\bibfield  {journal} {\bibinfo  {journal} {Journal of Computational
  Physics}\ }\textbf {\bibinfo {volume} {413}},\ \bibinfo {pages} {109451}
  (\bibinfo {year} {2020}{\natexlab{b}})}\BibitemShut {NoStop}%
\bibitem [{\citenamefont {Tzoufras}\ \emph {et~al.}(2008)\citenamefont
  {Tzoufras}, \citenamefont {Lu}, \citenamefont {Tsung}, \citenamefont {Huang},
  \citenamefont {Mori}, \citenamefont {Katsouleas}, \citenamefont {Vieira},
  \citenamefont {Fonseca},\ and\ \citenamefont {Silva}}]{Tzoufras2008}%
  \BibitemOpen
  \bibfield  {author} {\bibinfo {author} {\bibfnamefont {M.}~\bibnamefont
  {Tzoufras}}, \bibinfo {author} {\bibfnamefont {W.}~\bibnamefont {Lu}},
  \bibinfo {author} {\bibfnamefont {F.~S.}\ \bibnamefont {Tsung}}, \bibinfo
  {author} {\bibfnamefont {C.}~\bibnamefont {Huang}}, \bibinfo {author}
  {\bibfnamefont {W.~B.}\ \bibnamefont {Mori}}, \bibinfo {author}
  {\bibfnamefont {T.}~\bibnamefont {Katsouleas}}, \bibinfo {author}
  {\bibfnamefont {J.}~\bibnamefont {Vieira}}, \bibinfo {author} {\bibfnamefont
  {R.~A.}\ \bibnamefont {Fonseca}},\ and\ \bibinfo {author} {\bibfnamefont
  {L.~O.}\ \bibnamefont {Silva}},\ }\bibfield  {title} {\bibinfo {title} {{Beam
  loading in the nonlinear regime of plasma-based acceleration}},\ }\href@noop
  {} {\bibfield  {journal} {\bibinfo  {journal} {Physical Review Letters}\
  }\textbf {\bibinfo {volume} {101}},\ \bibinfo {pages} {145002} (\bibinfo
  {year} {2008})}\BibitemShut {NoStop}%
\bibitem [{\citenamefont {Dalichaouch}\ \emph {et~al.}(2021)\citenamefont
  {Dalichaouch}, \citenamefont {Xu}, \citenamefont {Tableman}, \citenamefont
  {Li}, \citenamefont {Tsung},\ and\ \citenamefont {Mori}}]{Dalichaouch2021}%
  \BibitemOpen
  \bibfield  {author} {\bibinfo {author} {\bibfnamefont {T.~N.}\ \bibnamefont
  {Dalichaouch}}, \bibinfo {author} {\bibfnamefont {X.~L.}\ \bibnamefont {Xu}},
  \bibinfo {author} {\bibfnamefont {A.}~\bibnamefont {Tableman}}, \bibinfo
  {author} {\bibfnamefont {F.}~\bibnamefont {Li}}, \bibinfo {author}
  {\bibfnamefont {F.~S.}\ \bibnamefont {Tsung}},\ and\ \bibinfo {author}
  {\bibfnamefont {W.~B.}\ \bibnamefont {Mori}},\ }\bibfield  {title} {\bibinfo
  {title} {{A multi-sheath model for highly nonlinear plasma wakefields}},\
  }\href@noop {} {\bibfield  {journal} {\bibinfo  {journal} {Physics of
  Plasmas}\ }\textbf {\bibinfo {volume} {28}},\ \bibinfo {pages} {063103}
  (\bibinfo {year} {2021})}\BibitemShut {NoStop}%
\bibitem [{\citenamefont {Yakimenko}\ \emph {et~al.}(2019)\citenamefont
  {Yakimenko}, \citenamefont {Alsberg}, \citenamefont {Bong}, \citenamefont
  {Bouchard}, \citenamefont {Clarke}, \citenamefont {Emma}, \citenamefont
  {Green}, \citenamefont {Hast}, \citenamefont {Hogan}, \citenamefont {Seabury}
  \emph {et~al.}}]{Yakimenko2019}%
  \BibitemOpen
  \bibfield  {author} {\bibinfo {author} {\bibfnamefont {V.}~\bibnamefont
  {Yakimenko}}, \bibinfo {author} {\bibfnamefont {L.}~\bibnamefont {Alsberg}},
  \bibinfo {author} {\bibfnamefont {E.}~\bibnamefont {Bong}}, \bibinfo {author}
  {\bibfnamefont {G.}~\bibnamefont {Bouchard}}, \bibinfo {author}
  {\bibfnamefont {C.}~\bibnamefont {Clarke}}, \bibinfo {author} {\bibfnamefont
  {C.}~\bibnamefont {Emma}}, \bibinfo {author} {\bibfnamefont {S.}~\bibnamefont
  {Green}}, \bibinfo {author} {\bibfnamefont {C.}~\bibnamefont {Hast}},
  \bibinfo {author} {\bibfnamefont {M.}~\bibnamefont {Hogan}}, \bibinfo
  {author} {\bibfnamefont {J.}~\bibnamefont {Seabury}}, \emph {et~al.},\
  }\bibfield  {title} {\bibinfo {title} {{FACET-II facility for advanced
  accelerator experimental tests}},\ }\href@noop {} {\bibfield  {journal}
  {\bibinfo  {journal} {Physical Review Accelerators and Beams}\ }\textbf
  {\bibinfo {volume} {22}},\ \bibinfo {pages} {101301} (\bibinfo {year}
  {2019})}\BibitemShut {NoStop}%
\bibitem [{\citenamefont {Freund}\ and\ \citenamefont
  {Antonsen}(1992)}]{Freund1992}%
  \BibitemOpen
  \bibfield  {author} {\bibinfo {author} {\bibfnamefont {H.~P.}\ \bibnamefont
  {Freund}}\ and\ \bibinfo {author} {\bibfnamefont {T.~M.}\ \bibnamefont
  {Antonsen}},\ }\href@noop {} {\emph {\bibinfo {title} {Principles of
  free-electron lasers}}}\ (\bibinfo  {publisher} {Springer},\ \bibinfo {year}
  {1992})\BibitemShut {NoStop}%
\end{thebibliography}%

\end{document}